\documentclass[pdflatex,sn-nature]{sn-jnl}

\usepackage{graphicx}%
\usepackage{multirow}%
\usepackage{amsmath,amssymb,amsfonts}%
\usepackage{amsthm}%
\usepackage{mathrsfs}%
\usepackage[title]{appendix}%
\usepackage{xcolor}%
\usepackage{textcomp}%
\usepackage{manyfoot}%
\usepackage{booktabs}%
\usepackage{algorithm}%
\usepackage{algorithmicx}%
\usepackage{algpseudocode}%
\usepackage{listings}%

\usepackage{xspace}
\usepackage[most]{tcolorbox}

\newcommand{\simulator}[0]{Light Society\xspace}

\theoremstyle{thmstyleone}%

\theoremstyle{thmstyletwo}%

\theoremstyle{thmstylethree}%

\begin{document}

\title[Modeling Earth-Scale Human-Like Societies with One Billion Agents]{Modeling Earth-Scale Human-Like Societies with One Billion Agents}

\author[1,2]{\fnm{Haoxiang} \sur{Guan}}
\equalcont{These authors contributed equally to this work.}

\author*[2,3]{\fnm{Jiyan} \sur{He}}\email{hejiyan@zgci.ac.cn}
\equalcont{These authors contributed equally to this work.}

\author[3]{\fnm{Liyang} \sur{Fan}}
\equalcont{These authors contributed equally to this work.}

\author[7,2]{\fnm{Zhenzhen} \sur{Ren}}

\author[2]{\fnm{Shaobin} \sur{He}}

\author[4]{\fnm{Xin} \sur{Yu}}

\author[5]{\fnm{Yuan} \sur{Chen}}

\author[2,3]{\fnm{Xueyin} \sur{Xu}}

\author[2,3]{\fnm{Shuxin} \sur{Zheng}}

\author[2,3]{\fnm{Yan} \sur{Gao}}

\author[1]{\fnm{Enhong} \sur{Chen}}

\author[2,3]{\fnm{Tie-Yan} \sur{Liu}}

\author*[2,3,6]{\fnm{Zhen} \sur{Liu}}\email{pt-lz@bza.edu.cn}

\affil[1]{\orgname{University of Science and Technology of China}, \orgaddress{\city{Hefei}, \country{China}}}

\affil[2]{\orgname{Zhongguancun Academy}, \orgaddress{\city{Beijing}, \country{China}}}

\affil[3]{\orgname{Zhongguancun Institute of Artificial Intelligence}, \orgaddress{\city{Beijing}, \country{China}}}

\affil[4]{\orgname{Shenzhen University}, \orgaddress{\city{Shenzhen}, \country{China}}}

\affil[5]{\orgname{Shanghai University of Finance and Economics}, \orgaddress{\city{Shanghai}, \country{China}}}

\affil[6]{\orgname{Tsinghua University}, \orgaddress{\city{Beijing}, \country{China}}}

\affil[7]{\orgname{Fudan University}, \orgaddress{\city{Shanghai}, \country{China}}}

\abstract{Understanding the dynamic evolution of complex social phenomena requires both high-fidelity modeling of human behavior and large-scale simulations. Traditional agent-based models (ABMs) have been employed to study these dynamics, but are constrained by simplified agent behaviors. Recent advances in large language models (LLMs) enable agents to exhibit sophisticated social behaviors, yet face significant scaling challenges.
We present \simulator, an agent-based simulation framework that advances both fronts. \simulator formalizes social processes as structured transitions of agent and environment states, governed by a set of LLM-powered simulation operations. Joint algorithmic and system optimizations, particularly a mixture-of-models engine that combines full LLMs with distilled surrogates, enable \simulator to efficiently simulate societies with over one billion agents.
Grounded in real-world demographic profiles from the World Values Survey, simulations of Trust Games and opinion diffusion at up to one billion agents demonstrate \simulator's high fidelity and efficiency in modeling diverse social phenomena, providing researchers with a practical foundation for hypothesis testing and the study of emergent collective behaviors at planetary scale.}

\keywords{Large Language Models, Agent-Based Modeling, Social Simulation, Emergent Behavior, Scalable Architectures}

\maketitle

\clearpage

\section{Main}
\label{sec:main}

The past decade has witnessed remarkable advancements in artificial intelligence, culminating in large language models (LLMs) that not only demonstrate unprecedented capabilities in language understanding and generation~\citep{openai2024gpt4technicalreport,brown2020languagemodelsfewshotlearners,zhao2025surveylargelanguagemodels}, but also achieve near-human performance on complex tasks requiring advanced reasoning including mathematics and coding~\citep{deepseekai2025deepseekr1incentivizingreasoningcapability,openai2025competitiveprogramminglargereasoning}. These capabilities have catalyzed the emergence of LLM-powered autonomous agents that are capable of perceiving their environment, making decisions, and taking actions in complex, dynamic settings~\citep{xi2023risepotentiallargelanguage,Wang_2024,Boiko2023}. Crucially, such agents can emulate nuanced individual behaviors, including maintaining personality traits~\citep{ICLR2024_c48dcd60,miotto-etal-2022-gpt,chen2024personapersonalizationsurveyroleplaying}, exhibiting self-awareness~\citep{li2024ithinkiam,BLUM202312}, and expressing context-sensitive emotions~\citep{huang2024emotionallynumbempatheticevaluating,10.19066/COGSCI.2024.35.1.002}, thereby capturing the fluidity and depth of human-like reactions. With the ability to simulate realistic and coherent individual behaviors, LLM-powered agents offer a new tool for investigating cognition, personality, and decision-making across diverse and reproducible scenarios.

Social science, as the study of human behavior in social contexts, offers key insights into how societies function and evolve. Traditional methods, such as questionnaires~\citep{foddy1993constructing,behr2018introduction}, interviews~\citep{Knott2022,roulston2003learning}, and controlled experiments~\citep{andreoni2021predicting,dehue2001establishing}, have long provided valuable insights, yet they are often costly, ethically constrained, and difficult to scale or replicate. Agent-based modeling (ABM) emerged as a computational alternative, enabling in silico experiments on social phenomena by simulating interactions among rule-based agents~\citep{axtell2025agent,axtell2016120,aylett2021june,aylett2022epidemiological,ghaffarian2021agent}. However, most ABMs rely on heuristics or fixed equations, limiting their capacity to reflect the complexity and adaptability of real-world societies~\citep{bonabeau2002agent}. The rise of large language models offers a transformative opportunity: agents powered by LLMs can generate context-rich, expressive, and adaptive behaviors, enabling simulations that integrate cognitive, emotional, and social dimensions. Recent work has demonstrated the promise of such agents in reproducing social dynamics like herd behavior~\citep{liu2024lmagent,yang2024oasis}, cooperation~\citep{guo2024embodied,vallinder2024cultural}, polarization~\citep{yang2024oasis,piao2025agentsociety}, information spreading~\citep{piao2025agentsociety,yang2024oasis}, presidential election~\citep{zhang2025socioverse,yu2024will}, and the impact of disease outbreaks~\citep{chopra2024limits}, suggesting that LLM-powered simulation may bridge longstanding gaps between computational models and the fluid realities of human societies.

Despite recent advances, current LLM-powered simulation frameworks remain prohibitively costly. For instance, many existing studies require dozens of GPUs and weeks of computation to simulate up to a million agents in social scenarios~\citep{yang2024oasis,piao2025agentsociety,liu2024lmagent,zhang2025socioverse,wang2025yulanonesimgenerationsocialsimulator,gao2024agentscopeflexiblerobustmultiagent,tang2024gensimgeneralsocialsimulation}. As simulations scale up, storage demands, latency, and overall computational costs grow rapidly, often rendering planet-scale simulations infeasible, with prior systems reaching at most $10^7$ agents (Supplementary Figure~\ref{fig:supp_scale_comparison}). Consequently, when attempting to study complex global phenomena, particularly those involving dynamic evolution and rich interactions that benefit from larger populations, scaling simulations to a global level while maintaining high fidelity becomes challenging and expensive. In addition, many existing frameworks are task-specific and lack generalizability~\citep{yang2024oasis,liu2024lmagent,ren2024baseslargescalewebsearch,mou2024unveilingtruthfacilitatingchange,zhang2024largescaletimeawareagentssimulation}, limiting their applicability across diverse social simulation conditions.

Here we introduce \simulator, a modular framework that decomposes social simulation into a small set of LLM-powered operators acting on agent and environment states, dispatched through an event queue. A unified stack of algorithmic and system co-optimizations, in which knowledge-distilled surrogate models and a mixture-of-models engine cut inference cost while compressed billion-node graphs and aggregated event execution drive runtime throughput, lets \simulator scale to populations of over one billion agents without sacrificing behavioral fidelity. We validate \simulator across a broad range of social scenarios with agents instantiated from real-world demographic profiles from the World Values Survey~\citep{haerpfer2022world}, anchored by two flagship case studies: large-scale Trust Games uncover scaling laws under which systematic demographic effects sharpen and individual variability diminishes as populations grow, while influencer-influencee opinion diffusion across a one-billion-agent network captures how opinions diffuse across the social network. By reducing the cost of population-scale, high-fidelity social experiments by orders of magnitude, \simulator offers the social sciences a reproducible testbed for hypothesis testing, counterfactual experimentation, and emergent-behavior research.

\section{Results}
\label{sec:results}

\subsection{Overview of \simulator}
\label{sec:overview_simulator}

\begin{figure}[t!]
    \centering
    \includegraphics[width=\textwidth]{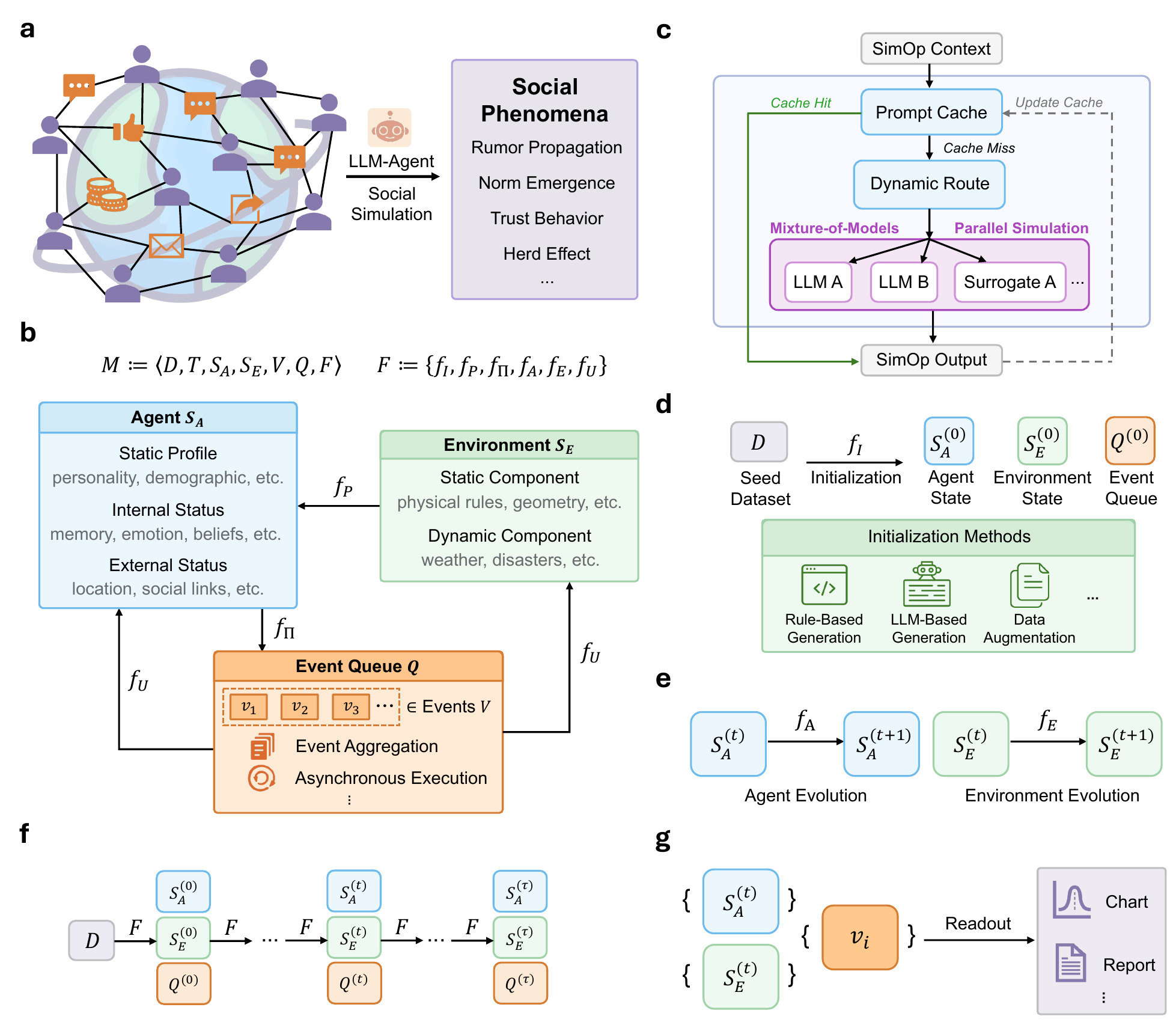}
    \caption{\textbf{Agent-based social simulation framework powered by LLMs.}
\textbf{a}, Conceptual overview of the social simulation platform capable of modeling various social phenomena including rumor propagation, norm emergence, trust behavior, and herd effects.
\textbf{b}, Formal specification of the simulation system, consisting of agents, environment, event queue, and simulation operations. Each agent maintains a static profile, internal status, and external status, while the environment encompasses both static and dynamic components.
\textbf{c}, Execution pipeline for simulation operations, using a mixture-of-models architecture with prompt caching and automatic routing across heterogeneous inference backends.
\textbf{d}, Initialization process using a seed dataset to generate the initial agent states, environment states, and event queue, with multiple initialization methods including rule-based generation, LLM-based generation, and data augmentation.
\textbf{e}, Autonomous evolution of agents and environment, capturing spontaneous changes such as memory decay, aging, and exogenous environmental dynamics independent of explicit interactions.
\textbf{f}, Full simulation process from seed dataset to termination under LLM-powered simulation operations.
\textbf{g}, Readout mechanism extracts structured outputs from the evolving agent and environment states along with events, enabling downstream analysis.}
    \label{fig:fig1_main}
\end{figure}

\simulator is organized around a single abstraction: every social interaction, whether between two agents in an economic game or across a billion-node network, is expressed as an LLM-powered operation that updates agent or environment state and is dispatched through an event queue (Figure~\ref{fig:fig1_main}). This uniform representation lets a single system host phenomena ranging from interpersonal trust to global opinion dynamics.

Three core entities compose \simulator: a population of agents, an environment, and an event queue that schedules all interactions in time (Fig.~\ref{fig:fig1_main}a,b). Each agent carries a static profile of persistent traits, an internal status of evolving cognition such as memory, beliefs, and goals, and an external status of observable features such as location and social connections. The environment combines static components such as spatial layout with dynamic components that drift as the simulation unfolds. Every interaction is encoded as an event, time-stamped and dispatched through the priority queue so that concurrent dynamics resolve deterministically. A small family of LLM-powered simulation operations drives all state changes among agents, the environment, and the event queue.

Scaling this design to a billion agents requires making both LLM inference and runtime communication efficient enough for billion-event throughput, which \simulator addresses through a jointly designed optimization stack (Fig.~\ref{fig:fig1_main}c). At the algorithmic layer, prompt caching reuses responses across repeated queries, lightweight surrogate models trained by knowledge distillation substitute LLM invocations for routine decisions, and a mixture-of-models inference engine routes each operation through a configurable policy, across backends ranging from full-scale LLMs for complex reasoning to task-specific surrogates for high-throughput updates. At the systems layer, compressed graph representations and vectorized batch operations make state updates and neighbor queries feasible on billion-node networks, while the event queue aggregates simultaneous events of the same type and dispatches them asynchronously, sharply reducing inter-agent latency and unlocking massive concurrency. Each component is exposed as an independently optimized module, allowing new accelerations or domain-specific primitives to be plugged in without altering the core simulation loop.

A complete simulation in \simulator proceeds through an initialization-evolution-readout cycle (Fig.~\ref{fig:fig1_main}d-g). An initialization operation generates the starting world from a domain-specific dataset, with various initialization methods including rule-based generation~\citep{chopra2024limits}, LLM-based generation~\citep{xie2024can}, and data augmentation~\citep{yang2024oasis}. As events fire, perception and policy operations let agents observe their surroundings and choose actions that emit new events, while agent and environment evolution operations capture autonomous changes between events, such as memory decay or environmental drift. An update operation applies pending events back to the global state, ensuring consistent transitions. After termination, a readout module aggregates the trajectory of system states and events into structured outputs suitable for downstream analysis.

\subsection{Trust Games}
\label{sec:trust_games}

To put \simulator into use, we begin with the Trust Game and evaluate how individual trust behavior and reciprocity vary across socio-economic backgrounds in large-scale LLM-powered simulations. In the Trust Game, a trustor decides how much of an initial \$10 endowment to send to an anonymous trustee, with the amount tripled upon transfer~\citep{cox2004identify,berg1995trust,xie2024can}. The trustee then decides how much of the received amount to return. This minimal setup captures key dynamics of social exchange, including risk-taking, fairness, and reciprocity. We also implemented and evaluated several Trust Game variants (collectively referred to as \textit{Trust Games}) to more broadly explore patterns of human trust behavior and assess the capabilities of \simulator in modeling diverse social exchange scenarios (see Section~\ref{sec:trust_games_setup} and Appendix~\ref{app:trust_variants_results}).

\begin{figure}[t!]
    \centering
    \includegraphics[width=\textwidth]{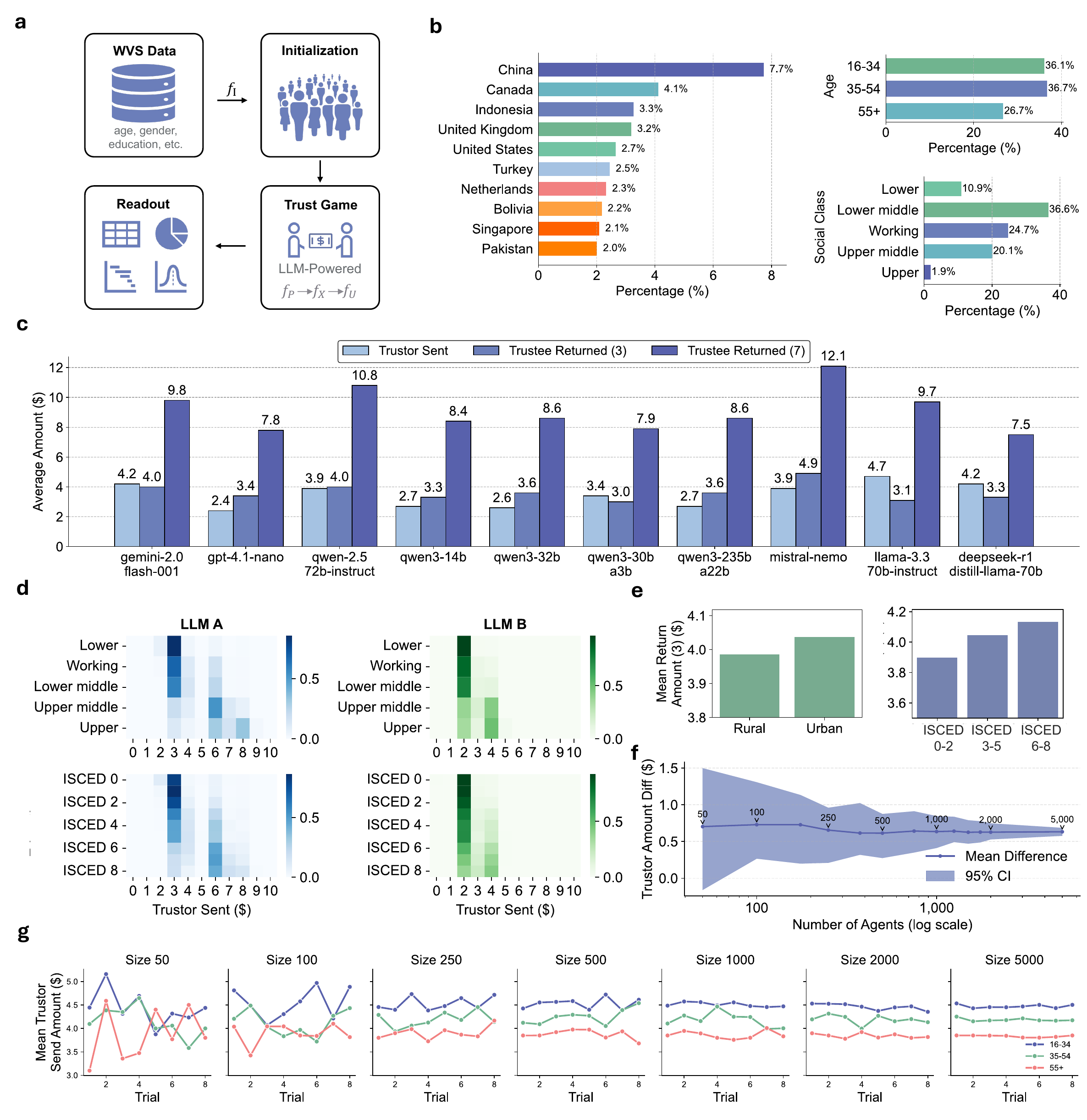}
    \caption{\textbf{Trust Game simulation.}
\textbf{a}, Overview of the simulation pipeline.
\textbf{b}, Demographic distribution of the simulated agent population.
\textbf{c}, Average trustor sending amount and trustee returning amount when receiving \$3 or \$7 across diverse LLMs.
\textbf{d}, Trustor behavior stratified by subjective social class and education level across two LLMs (A: \texttt{gemini-2.0-flash-001}, B: \texttt{gpt-4.1-nano}).
\textbf{e}, Average trustee return amount when trustor sending \$3, stratified by urban-rural residence and education level.
\textbf{f}, Scaling behavior of the difference in trustor sending amounts between the 16–34 and 55+ age groups.
\textbf{g}, Average trustor sending amount by age group over 7 simulation trials across different population sizes. Trust levels stabilize as population scales up, with younger agents consistently sending more.}
    \label{fig:fig2_trust}
\end{figure}

Our simulation pipeline (Figure~\ref{fig:fig2_trust}a) uses real-world profiles constructed from the World Values Survey (WVS) Wave 7~\citep{haerpfer2022world}, which offers extensive coverage of individual characteristics across a wide range of socio-demographic and cultural dimensions, including country, age, and subjective social class (Figure~\ref{fig:fig2_trust}b). Each agent is instantiated with a natural language profile derived from real-world survey responses, and both the trustor and trustee roles are played by LLMs, conditioned on these profiles.

We first analyzed how trustor decisions vary with socio-economic status. Figure~\ref{fig:fig2_trust}d shows the distribution of send amounts stratified by subjective social class and education level. Across both models, trust behavior exhibited clear upward trends with increasing social class and education attainment. Agents identifying as upper class were substantially more likely to send higher amounts compared to those from lower classes, and similarly, individuals with advanced education levels (e.g., postgraduate) demonstrated higher trust than those with little or no formal education. These stratified patterns were evident across both models, suggesting that socio-economic background may influence trust propensity in systematic ways.

While the relative trends remain consistent across socio-demographic groups, we observed clear differences in the overall level of trust expressed by agents simulated with different models (Figures~\ref{fig:fig2_trust}c,d). Average trustor send amounts and trustee return amounts varied across models, indicating that model-specific inductive biases influence both parties' decision patterns. Nevertheless, the demographic-driven trends remained stable across models, suggesting that \simulator captures robust and interpretable social patterns despite model-level variability, thereby providing a reliable basis for studying social behavior at scale.

Trustee behavior revealed structured patterns of reciprocity that varied with the amount received (simulated by \texttt{gemini-2.0-flash-001}). As shown in Supplementary Figure~\ref{fig:supp_trust}a, average return amounts increased approximately linearly with trustor send amounts, while return ratios fluctuated across transfer levels. Trustors achieved positive net profits across most transfer levels, indicating that trust was generally rewarded. This emergent pattern suggests that LLM agents simulate reciprocal norms that support mutually beneficial exchange, hallmarks of prosocial behavior in human societies. Further stratification at a fixed transfer level (Figure~\ref{fig:fig2_trust}e) revealed that reciprocity also varied with residence and education level, suggesting that social advantage may be associated with stronger tendencies toward reciprocal behavior.

We next investigated scaling laws in LLM-powered social simulations by varying the number of agents participating in the Trust Game, simulated by \texttt{gemini-2.0-flash-001}. As shown in Figure~\ref{fig:fig2_trust}f, the gap in sending amounts between younger (16–34) and older (55+) trustors becomes increasingly pronounced as population size grows, with confidence intervals narrowing accordingly. This indicates that larger populations reduce stochastic fluctuations in individual decision-making and amplify systematic demographic effects. The stability of this effect was further validated across seven independent trials for each population size (Figure~\ref{fig:fig2_trust}g), where trustor behaviors converged toward stable equilibria and age-related differences remained consistent. These results highlight that large-scale simulations not only reproduce known behavioral regularities but also uncover latent social patterns that are difficult to detect in smaller populations, providing a powerful lens for examining how demographic factors and social norms interact under conditions of scale.

\subsection{Opinion Diffusion via Influencer-Influencee Interactions with One Billion Agents}
\label{sec:1b}

\begin{figure}[t!]
    \centering
    \includegraphics[width=\textwidth]{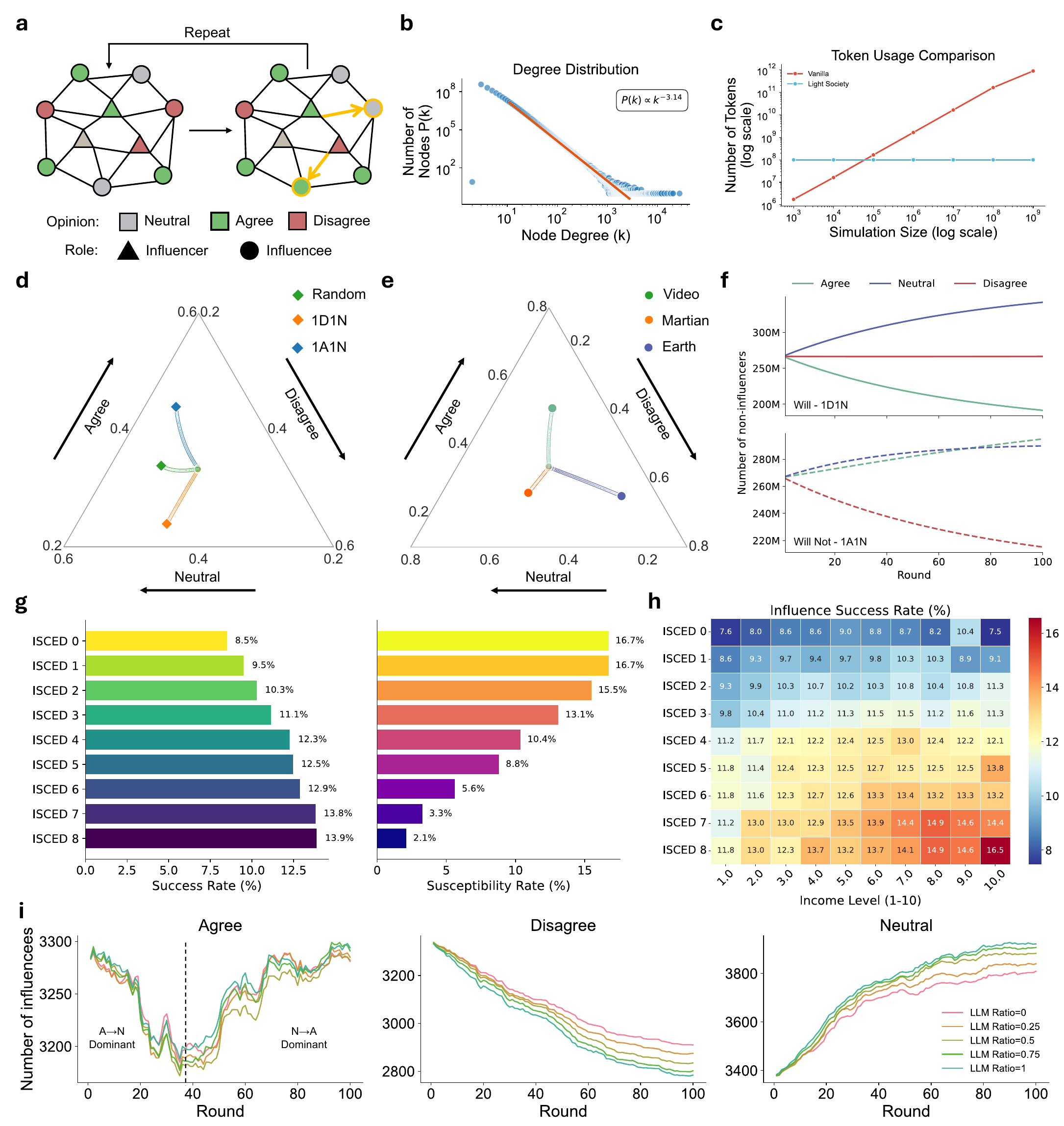}
    \caption{\textbf{Opinion diffusion via influencer-influencee interactions across a one-billion-agent network.}
\textbf{a}, Conceptual overview of the simulation pipeline.
\textbf{b}, Degree distribution of the generated BA network.
\textbf{c}, Cumulative LLM token usage over simulation size, comparing full-LLM and surrogate-assisted inference modes.
\textbf{d}, Evolution of influencee opinion distributions over simulation rounds under each seeding scheme (1A1N, 1D1N, Random), showing distinct opinion trajectories for a single topic.
\textbf{e}, Opinion trajectories under a fixed seeding scheme (1A1N) across different topics, demonstrating topic-dependent convergence patterns.
\textbf{f}, Opinion shift curves for dual configurations: ``AI automation will lead to mass unemployment'' with 1D1N seeding versus its negation ``AI automation will not lead to mass unemployment'' with 1A1N seeding. The mirrored setups yield consistent directional trends with quantitative differences.
\textbf{g}, Influence success rates by influencer education level and resistance to influence by influencee education level, showing asymmetric persuasion dynamics.
\textbf{h}, Joint effect of education and income on influence success rate; higher socioeconomic status leads to greater influence success rate.
\textbf{i}, Validation of surrogate model fidelity in a 10,000-agent network: opinion evolution trajectories under varying surrogate substitution levels (0\%, 25\%, 50\%, 75\%, 100\%) remain qualitatively consistent with the full-LLM baseline, confirming that the surrogate preserves behavioral dynamics while substantially reducing computational cost.}
    \label{fig:1b}
\end{figure}

Opinion dynamics explores how individual attitudes evolve through social interactions, shaping collective beliefs across populations~\citep{xia2011opinion,das2014modeling}. Many real-world networks, including social and communication networks, can be approximated as scale-free~\citep{barabasi1999emergence}, in which a small subset of highly connected ``influencers'' wield disproportionate impact on opinion formation~\citep{rochert2022two,mohamed2012identifying,brown2008influencer}. To push \simulator to planetary scale, we simulate opinion diffusion over a one-billion-agent Barabási–Albert (BA) network (Fig.~\ref{fig:1b}a,b). Each agent is assigned a demographic profile sampled from a pool of 10,000 profiles drawn from the World Values Survey (WVS) Wave 7 dataset~\citep{haerpfer2022world}. The top 20\% of nodes by degree serve as \textit{influencers}; the remaining 80\% are \textit{influencees}. All agents are given a controversial statement, ``AI automation will lead to mass unemployment'', and assigned an initial opinion (\texttt{agree}, \texttt{disagree}, or \texttt{neutral}). Influencee opinions are initialized randomly, while influencer opinions are configured under three seeding schemes: \textbf{1D1N} (equal \texttt{disagree}/\texttt{neutral}), \textbf{1A1N} (equal \texttt{agree}/\texttt{neutral}), and \textbf{Random}. In each round, 1\% of influencers interact with their immediate influencees, potentially updating the influencee's stance.

For tractable inference at this scale, we distill LLM behaviors into a lightweight MLP surrogate model trained on interaction samples spanning diverse profile and opinion combinations. The surrogate replaces most LLM calls, substantially reducing token usage (Fig.~\ref{fig:1b}c). A detailed evaluation of surrogate fidelity is presented in Section~\ref{sec:analysis}.

As shown in Fig.~\ref{fig:1b}d, the initial seeding of influencer opinions critically shapes population-level attitude evolution. Under \textbf{1A1N}, the opinion centroid shifts markedly toward \texttt{agree}; under \textbf{1D1N}, the population drifts toward \texttt{neutral} rather than \texttt{disagree}; and under \textbf{Random}, the trajectory falls between the two with a mild residual drift toward \texttt{agree}. This asymmetry indicates that the influencee population carries an inherent prior favoring the statement, so that dissent-seeded influence primarily induces neutralization rather than reverse persuasion.

Beyond this single statement, we apply the \textbf{1A1N} scheme to three additional topics (Fig.~\ref{fig:1b}e): ``The Earth is flat'' (strong scientific counter-prior), ``Humans will establish a Martian city within 50 years'' (high uncertainty), and ``Short-form videos are reducing human attention spans'' (broad societal resonance). The resulting trajectories diverge in ways tightly coupled with each topic's prior: the flat-Earth centroid migrates toward \texttt{disagree}, the Martian-city centroid drifts toward \texttt{neutral}, and the short-videos centroid shifts toward \texttt{agree}. These results demonstrate that influencer skew does not mechanically steer populations in arbitrary directions; rather, its effect is jointly modulated by the prior distribution, social consensus, and topic plausibility. The stance transition distributions learned by the surrogate model for these three topics are compared with the ground-truth LLM outputs in Supplementary Figures~\ref{fig:supp_trans_earth},~\ref{fig:supp_trans_martian},~\ref{fig:supp_trans_video}.

We next analyze how demographic attributes shape influence dynamics. As shown in Fig.~\ref{fig:1b}g, agents with higher education levels are more successful at persuading others while also being more resistant to opinion change themselves. Fig.~\ref{fig:1b}h extends this analysis to the joint effect of education and income: influence success rate increases with both dimensions, with the highest efficacy among agents with postgraduate education and high income.

Linguistic framing introduces a separate axis: we compare two dual configurations (Fig.~\ref{fig:1b}f): ``AI will cause mass unemployment'' with \textbf{1D1N} seeding versus ``AI will not cause mass unemployment'' with \textbf{1A1N} seeding. Although both configurations push the population away from the same underlying belief, the negative framing primarily induces neutralization while the positive framing induces more substantive stance change, consistent with the schema-plus-tag account of negation processing, in which negated propositions retain the core affirmative schema while the negation marker is readily lost during downstream cognition~\citep{mayo2004not,hasson2006negation}.

Finally, we validate surrogate fidelity by comparing opinion evolution trajectories in a 10,000-agent network under five surrogate substitution levels (0\%, 25\%, 50\%, 75\%, 100\%). As shown in Fig.~\ref{fig:1b}i, trajectories remain qualitatively consistent across all levels, with near-identical inflection points and convergence behavior, confirming that the surrogate preserves the behavioral dynamics of the full LLM while enabling simulation at billion-agent scale.

Notably, all trajectories in Fig.~\ref{fig:1b}i exhibit a U-shaped agree curve, in which the agree fraction first declines before recovering. Further analysis reveals that direct transitions between agree and disagree are rare; instead, opinion change proceeds almost exclusively through the neutral state, which serves as a transitional buffer. The initial decline is consistent with psychological reactance, whereby agents resist external influence attempts by retreating to a noncommittal stance~\citep{brehm1966reactance,knowles2004resistance}, while the subsequent recovery aligns with informational cascade dynamics, in which accumulated social proof from converted neighbors progressively triggers acceptance among remaining neutral agents once a local threshold is reached~\citep{bikhchandani1992cascade,granovetter1978threshold,centola2010spread}.

\subsection{Analysis}
\label{sec:analysis}

We conduct a series of analyses to further examine the behavioral richness, robustness, and generality of \simulator. Specifically, we investigate variance and stability in the influencer-influencee opinion dynamics, long-horizon simulation, the effect of communication language, sensitivity to surrogate model choice, the role of social network topology, game-theoretic validation of human-like bounded rationality, and memory-augmented free discussion under less constrained interaction.

\subsubsection{Variance and Long-Horizon Dynamics}

The influencer-influencee simulations performed with \simulator carry two natural sources of stochasticity: the random sampling of active influencers in each round, and the intrinsic non-determinism of LLM inference when live LLM calls are used in place of the surrogate. We isolate each source in turn by holding one fixed and allowing the other to vary, and report variability using the coefficient of variation (CV), defined as the ratio of the across-run standard deviation to the across-run mean at each round, which gives a scale-free measure of run-to-run dispersion.

We first isolate variability from per-round influencer sampling by fixing the topic (``The Earth is flat'') and the 1A1N seeding scheme, use the embedding-MLP surrogate exclusively so that LLM noise is removed, and run the full billion-agent simulation five times with different random seeds. The resulting per-round opinion trajectories (top) and across-run standard deviations (bottom) for the agree, neutral, and disagree populations are shown in Fig.~\ref{fig:analysis}a (left). Across the 100-round horizon, the peak CV reaches only 0.0038\% for agree, 0.0060\% for neutral, and 0.0038\% for disagree, meaning that the spread across the five runs never exceeds a ten-thousandth of the corresponding mean, so per-round influencer sampling contributes negligible run-to-run dispersion at the billion-agent scale. We then isolate variability from LLM sampling by fixing the random seed and rerun a mixed-inference variant five times on a one-million-agent network under the same 1A1N, Earth-is-flat configuration, resolving 50\% of interactions through live LLM calls and the remaining 50\% through the surrogate. Because the seed is held constant, any residual run-to-run spread reflects the non-determinism of the LLM. The corresponding mean trajectories and standard deviation bands are plotted in Fig.~\ref{fig:analysis}a (right). Peak CVs rise to 0.1448\% for agree, 0.3365\% for neutral, and 0.1616\% for disagree, about two orders of magnitude larger than in the pure-surrogate case but still safely below half a percent throughout the simulation, indicating that population-level dynamics remain robust to the additional noise introduced by live LLM inference.

On a longer horizon, we extend the simulation to 5{,}000 rounds for all nine combinations of three topics and three seeding schemes (Supplementary Fig.~\ref{fig:supp_long_horizon}). Every configuration eventually settles into a dynamic equilibrium in which opinion fractions fluctuate within a narrow band rather than collapsing onto a single stance, consistent with the view from statistical social dynamics that opinion systems generically relax into stochastic stationary states with persistent heterogeneity rather than full consensus~\citep{castellano2009statistical}. The location of this equilibrium is jointly shaped by the topic and the seeding scheme.

\subsubsection{Effect of Communication Language}

Because \simulator realizes interactions through natural language, the choice of communication language may itself modulate the simulated dynamics. To quantify this, we regenerate the influencer-influencee interaction dataset with Gemini 2.0 Flash for three topics, ``The Earth is flat'', ``Humans will establish a Martian city within 50 years'', and ``Short-form videos are reducing human attention spans'', under two languages, Chinese and French, while keeping the WVS agent profiles, stance-pair conditions, and prompt template otherwise identical.

We summarize each combination by its stance change rate, defined as the fraction of interactions in which the influencee's final stance differs from the initial stance. As shown in Fig.~\ref{fig:analysis}b, the three topics occupy clearly separated change-rate levels, with Earth-is-flat near 45\%, short videos near 27\%, and Martian city near 22\%, reflecting topic-intrinsic volatility that is largely consistent with the priors identified in Section~\ref{sec:1b}. On top of this topic-level variation, swapping the communication language produces measurable shifts: the change rate is 45.22\% for Chinese and 47.25\% for French on Earth-is-flat, 22.53\% and 21.61\% on Martian city, and 30.05\% and 25.34\% on short videos. Both the magnitude and the sign of the language effect depend on the topic. The gap grows from under one percentage point on Martian city to roughly 4.7 percentage points on short videos, and while French elicits slightly more stance change than Chinese on Earth-is-flat, Chinese elicits more stance change than French on the other two topics. No single language is uniformly ``more persuasive''; rather, language interacts with the topic to shape how readily influencees update. This topic-contingent language effect parallels empirical findings in human studies that the language of communication itself alters judgment, such as the foreign-language effect on decision biases~\citep{keysar2012foreign} and the persistent influence of interview language on self-reported political opinion within otherwise comparable populations~\citep{lee2014persistent}.

To see whether these interaction-level differences propagate through a full diffusion simulation, we focus on the short-videos topic, which shows the largest Chinese / French gap, and run the influencer-influencee dynamics separately under each language for all three seeding schemes. The resulting opinion trajectories (Supplementary Fig.~\ref{fig:supp_lang_diff}) visibly diverge between the two languages under every seeding scheme, confirming that microscopic language-induced differences in stance updating are not washed out at the population level and instead translate into distinct macroscopic outcomes. Communication language therefore constitutes an additional factor, beyond agent profile, network topology, and seeding scheme, that \simulator-based studies should control for or explicitly compare when interpreting simulated societal dynamics.

\subsubsection{Sensitivity to Surrogate Model Choice}

The billion-agent influencer-influencee simulation in \simulator relies on a surrogate model trained to imitate the teacher LLM, so the choice of surrogate architecture can in principle shape the resulting dynamics. We train five surrogate architectures on the same teacher-LLM interaction samples: a softmax regression (SR) and a gradient-boosted tree ensemble (LightGBM~\citep{ke2017lightgbm}) operating on raw WVS demographic attributes, a multilayer perceptron (MLP) and a Transformer~\citep{vaswani2017attention} operating on pre-computed profile embeddings, and a Qwen3-0.6B~\citep{qwen3technicalreport} small language model (SLM) supervised-fine-tuned on the same teacher-LLM interaction samples using the same prompt template as the teacher LLM.

\begin{figure}[t!]
    \centering
    \includegraphics[width=\textwidth]{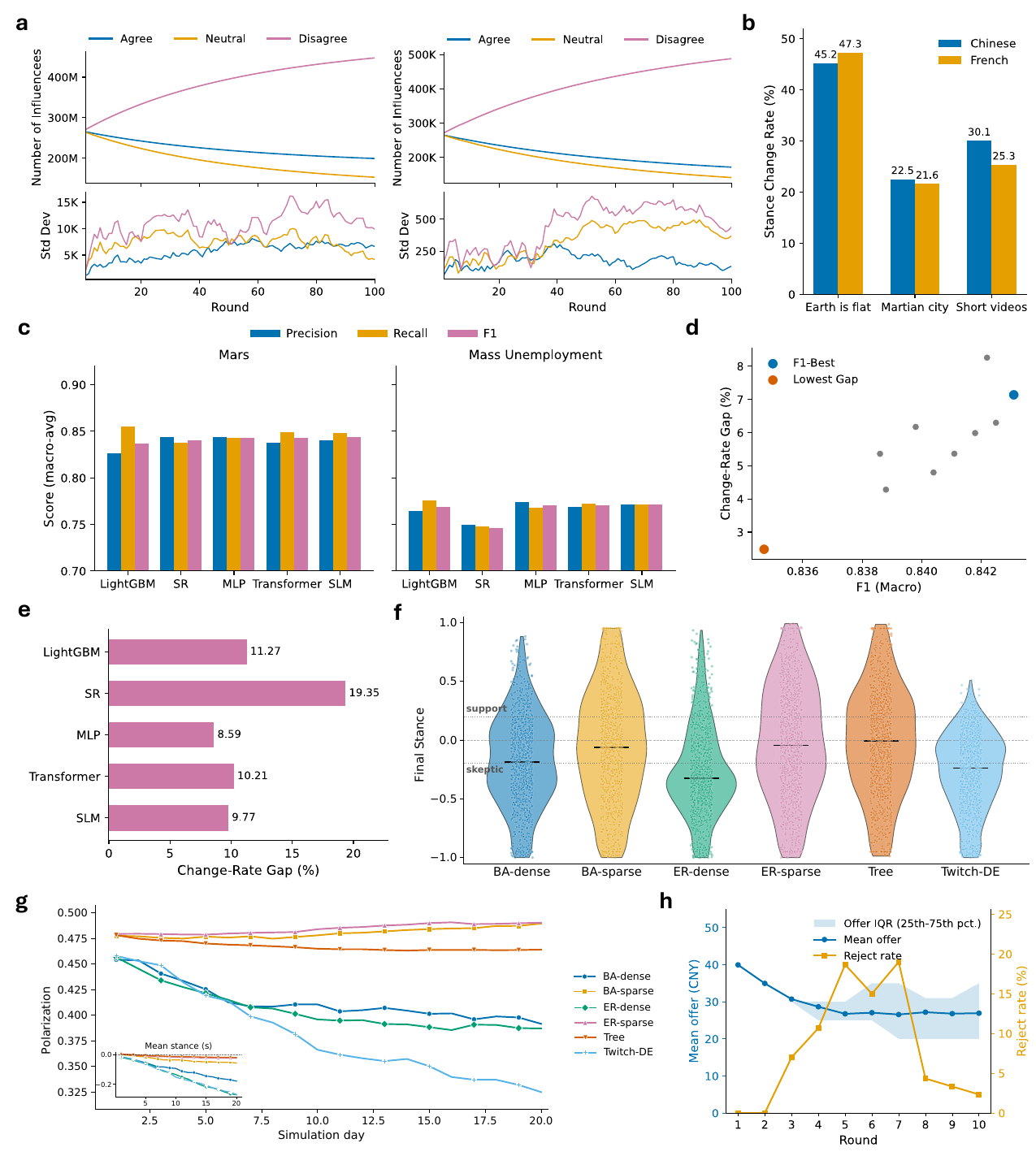}
    \caption{\textbf{Multi-faceted analyses of \simulator.}
\textbf{a}, Across-run variance of the influencer-influencee simulation on ``Earth is flat'' under 1A1N seeding. Left: billion-agent simulation, embedding-MLP surrogate, five seeds. Right: one-million-agent simulation, fixed seed, 50\%-LLM / 50\%-surrogate mixed inference, five runs. Top: mean trajectories; bottom: across-run std.
\textbf{b}, Stance change rate for three topics under Chinese and French communication languages.
\textbf{c}, Macro-averaged precision, recall, and F1 of the five surrogate architectures at their F1-best checkpoints, on the Martian-city (left) and Mass Unemployment (right) topics.
\textbf{d}, F1 versus change-rate gap across the ten training epochs of the Martian-city MLP, with the F1-best and lowest-gap epochs highlighted.
\textbf{e}, Change-rate gap at the F1-best checkpoint for the five surrogate architectures on the Mass Unemployment topic.
\textbf{f}, Day-20 per-agent stance distributions across six network substrates ($N = 1{,}000$): three sparse (BA $m=3$, ER $\langle k \rangle = 6$, random tree), two density-matched dense (BA $m=27$, ER $\langle k \rangle = 54$), and the empirical Twitch-DE subgraph. Dotted lines mark $s = \pm 0.2$.
\textbf{g}, Twenty-day trajectories of the stance standard deviation $\sigma(s)$ for the six substrates in \textbf{f}; inset, corresponding mean-stance $\langle s \rangle$ trajectories.
\textbf{h}, Round-by-round mean offer (left, with 25th--75th-percentile band) and rejection rate (right) in the 10-round Ultimatum Game.}
    \label{fig:analysis}
\end{figure}

At the per-sample classification level, the five surrogates are broadly comparable within each topic (Fig.~\ref{fig:analysis}c). On the Martian-city topic, macro F1 is tightly clustered between 0.84 and 0.85 across all five architectures; on the Mass Unemployment topic it lies in a narrower band between 0.75 and 0.77. The two topics occupy clearly separated F1 ranges, indicating that prediction difficulty is primarily driven by the topic rather than by the architecture, yet the within-topic model-level differences are small but non-negligible once aggregate fidelity is considered.

To capture aggregate distributional fidelity we additionally report the change-rate gap, defined as the absolute difference between the surrogate's fraction of stance-changing interactions and the teacher LLM's fraction. Within a single model, higher per-sample F1 does not imply a smaller change-rate gap. Tracing the Martian-city MLP over its ten training epochs (Fig.~\ref{fig:analysis}d), the F1-maximizing checkpoint (epoch 6, F1 $=$ 0.843, gap $=$ 7.14\%) is roughly three times worse in gap than the gap-minimizing checkpoint (epoch 9, F1 $=$ 0.835, gap $=$ 2.48\%), despite an F1 difference of only 0.008. A checkpoint with slightly lower per-sample accuracy can therefore yield substantially better aggregate fidelity, so high F1 alone is not a reliable indicator of surrogate quality.

The same pattern holds across architectures. Evaluating the five surrogates at their respective F1-best checkpoints on Mass Unemployment, the change-rate gap ranges from 8.6 percentage points for the MLP to 19.4 percentage points for SR (Fig.~\ref{fig:analysis}e). The spread across models on this distributional metric is much wider than their per-sample F1 spread would suggest, and the ordering of architectures is not aligned with the F1 ordering. Selecting a surrogate for deployment therefore benefits from combining multiple criteria. In practice, one can first shortlist a top-$N$ set of checkpoints by conventional training metrics such as F1, and then choose within this shortlist the checkpoint whose post-training aggregate distribution most closely matches that of the teacher LLM. On this basis, the embedding MLP is selected as the surrogate for the billion-agent simulation reported in Section~\ref{sec:1b}.

We finally note that, unlike the other four surrogates, the fine-tuned SLM is a generative model whose outputs are stochastic, just as the teacher LLM's are. On complex topics where the teacher's transition distribution is itself highly non-deterministic, a generative surrogate can in principle capture this distributional structure more faithfully than a deterministic classifier, making the SLM family a promising candidate when aggregate fidelity is the dominant concern.

\subsubsection{Effect of Social Network Topology}
\label{sec:effect_of_social_network_topology}

We next probe how the network substrate itself shapes collective opinion dynamics within \simulator, running an additional series of simulations at a fixed scale of $N = 1{,}000$ agents under a modified interactional setup. Each agent maintains a continuous stance $s \in [-1, +1]$ and confidence $c \in [0, 1]$, and any agent can act as either speaker or listener. The simulation is driven by a sequence of speak, heard, and reflective-update events dispatched through the event queue of \simulator, with all LLM-powered state updates executed by \texttt{deepseek-v4-flash}. We compare six network substrates on the same set of 1{,}000 agents: three sparse synthetic baselines (Barab\'{a}si--Albert, Erd\H{o}s--R\'{e}nyi, and a random tree), two density-matched dense synthetic graphs, and one empirical substrate, a connected 1{,}000-node subgraph of the Twitch-DE friendship network from the MUSAE dataset~\citep{rozemberczki2021multi}. Full setup details are given in Section~\ref{sec:network_topology_setup}.

Over 20 simulation days, the six trajectories of population-level opinion spread $\sigma(s)$, defined as the standard deviation of per-agent stance, separate into two clearly distinct groups (Fig.~\ref{fig:analysis}g). The three sparse graphs keep $\sigma(s)$ flat between 0.46 and 0.49 with a near-zero mean stance $\langle s \rangle$, whereas the three dense graphs, including Twitch-DE, show a monotonic decline in $\sigma(s)$ to between 0.33 and 0.39 accompanied by a drift of $\langle s \rangle$ down to between $-0.18$ and $-0.27$. The final stance distributions (Fig.~\ref{fig:analysis}f) corroborate this grouping: the sparse graphs produce broad, symmetric distributions with support and skepticism fractions close to the random initialization (33\%--39\% each), while the dense graphs yield compressed, downward-shifted distributions in which the skepticism fraction rises to 49\%--64\% and the support fraction falls to 5\%--16\%. Per-topology summary statistics are tabulated in Supplementary Fig.~\ref{fig:supp_network_table}.

Within the dense group, the empirical Twitch-DE network lies close to the two density-matched synthetic graphs rather than forming a distinct category. Its polarization amplitude $|\langle s \rangle| = 0.269$ is essentially indistinguishable from that of the density-matched Erd\H{o}s--R\'{e}nyi graph (0.270), and its skepticism fraction (54\%) is even lower than the latter's (64\%), despite Twitch-DE's substantially higher clustering coefficient ($C = 0.31$ versus $0.05$). Aligning the mean degree of a simple Erd\H{o}s--R\'{e}nyi graph with that of the empirical network closes roughly 57\% of the polarization gap between the sparse synthetic baselines and Twitch-DE, reducing $|\Delta \langle s \rangle|$ from 0.21 to 0.09.

\subsubsection{Human-like Bounded Rationality}

Bounded rationality is a hallmark of human decision-making; we test whether the LLM agents underlying \simulator reproduce it by running the classic Ultimatum Game, in which the game-theoretic equilibrium (the proposer offers a minimal amount and the responder accepts any positive amount) is known to diverge sharply from observed human behavior. We use a single-round reference condition together with a 10-round iterated setup involving 300 proposer-responder pairs and 3{,}000 pair-round interactions in total; agent profiles are drawn from the same WVS pool used elsewhere in this section.

In the single-round reference condition, the mean offer is 41.0 out of 100, squarely within the 40\%--50\% range commonly reported in human experiments~\citep{guth1982experimental, henrich2001search} and far from the near-zero offer predicted by pure rationality. The aggregate offer distribution and offer-conditioned rejection risk are shown in Supplementary Fig.~\ref{fig:supp_ultimatum}: low offers face sharply elevated rejection rates, indicating that agents willingly forgo material gain to punish inequitable proposals.

The 10-round iterated setup reveals additional adaptive dynamics (Fig.~\ref{fig:analysis}h). The mean offer declines from 40.0 in the first round to 26.9 in the tenth, while the rejection rate climbs through the middle rounds, peaks at 19.0\% in round 7, and falls back to around 2\% by the end. This trajectory is consistent with patterns reported in repeated-ultimatum experiments with humans, and the systematic deviation from game-theoretic rationality shows that LLM agents in \simulator capture the non-rational components of human bargaining behavior that rule-based ABMs cannot.

\subsubsection{Memory-Augmented Free Discussion}

Beyond the influencer-influencee setup, we test whether richer memory and unrestricted communication alter the collective dynamics captured by \simulator, running a memory-augmented free-discussion simulation with 150 agents sampled from WVS demographic profiles. Any pair of agents can converse, and each agent maintains a sliding window of recent conversations together with daily reflective summaries that feed into subsequent behavior. The topic is whether humanity can establish a Martian city within 50 years.

At the population level, the centroids of the two initial opinion clusters move toward a shared region in semantic embedding space over seven days (Supplementary Fig.~\ref{fig:supp_free_discuss}a), and the overall cosine diversity of the population drops from 0.196 to 0.125, a 36\% reduction. Memory-augmented open discussion therefore still drives the population toward a shared stance even without centralized influencers or interaction constraints.

Convergence at the population level does not, however, imply homogenization of individual reasoning. The mean within-cluster spread, measured as distance from each cluster's centroid, increases from Day 0 to Day 7 by 3.1$\times$ in one cluster and 2.1$\times$ in the other (Supplementary Fig.~\ref{fig:supp_free_discuss}b). Agents become more similar in their high-level conclusions while developing more varied rationales within each stance, which shows that \simulator captures heterogeneous reasoning structures alongside macro-level convergence under enriched memory and unrestricted social interaction.

\section{Discussion}
\label{sec:discussion}

By scaling LLM-powered social simulation to a billion agents while preserving cognitively rich, demographically grounded behavior, \simulator narrows the gap between in-silico social experiments and the population-scale phenomena they are meant to model. The Trust Game and one-billion-agent opinion-diffusion case studies, together with the analyses of variance, language, surrogate choice, network topology, bounded rationality, and free discussion, demonstrate that this scale is reachable in practice and produces interpretable, reproducible social dynamics.

Beyond engineering convenience, planet-scale simulation has scientific payoff. The Trust Game scaling experiments show that systematic demographic effects, such as the gap in trust behavior between younger and older trustors, sharpen as the population grows, while stochastic fluctuations in individual decisions diminish (Fig.~\ref{fig:fig2_trust}f,g). At the small populations typical of prior LLM-powered simulations, such effects can be drowned in individual noise; at billion-agent scale they become detectable and reproducible. By making this regime computationally tractable via mixture-of-models inference and distilled surrogates, \simulator opens for social scientists a class of population-scale, hypothesis-driven experiments that were previously infeasible to run in silico.

Once at this scale, \simulator's outputs are highly reproducible. Across five independent runs of the billion-agent influencer-influencee simulation, the per-round coefficient of variation in opinion fractions stays below 0.01\%, confirming that conclusions drawn from a single run are not artefacts of stochastic seeding (Fig.~\ref{fig:analysis}a). Yet variability is not always a defect: the richer source of variation lies elsewhere, across LLMs, communication languages, surrogate architectures, and network topologies, where the same experimental setup can produce measurably different macroscopic outcomes. Rather than treating these axes as nuisance variables to be averaged out, researchers can use them as parameters of the experimental design, mapping how robust any specific finding is to the modeling choices that produced it.

This robustness is, however, conditional on the underlying language model. Trust Games show that absolute send and return amounts vary across LLMs even when socio-economic gradients remain stable (Fig.~\ref{fig:fig2_trust}c,d), and the communication-language analysis shows that swapping Chinese for French can shift the stance-change rate by up to 4.7 percentage points with a topic-dependent sign. Quantitative claims drawn from a single model and language should therefore be interpreted as one realization in a family of LLM-mediated social dynamics; studies built on \simulator should explicitly compare across these choices where conclusions are sensitive to them, rather than treating any one configuration as a neutral oracle.

On the methodological side, the surrogate-sensitivity analysis carries a practical lesson: per-sample classification accuracy alone is not a reliable indicator of surrogate quality at the population level, because checkpoints with similar F1 can differ several-fold in their distributional gap from the teacher LLM (Fig.~\ref{fig:analysis}d,e). Selecting a deployable surrogate benefits from combining per-sample and aggregate-distribution criteria, and generative small language models, fine-tuned on teacher interactions, are a promising candidate when distributional fidelity is the dominant concern.

\simulator's reach also raises ethical considerations that we believe should be addressed in tandem with capability. A planet-scale, controllable simulator of opinion diffusion lowers the barrier not only to legitimate research but also, in principle, to the design of large-scale information operations and persuasive content; we therefore intend the released framework as an instrument for studying such dynamics, not for executing them. The framing of agent populations also inherits the demographic biases of both the World Values Survey and the underlying LLM, which can over- or under-represent specific groups in ways that are not always visible in aggregate statistics. Finally, simulated populations should not be treated as substitutes for human participants: \simulator's outputs are best read as hypotheses to be examined empirically, not as evidence about real societies on their own.

The same operator-and-event abstraction extends naturally beyond the social-science setting, to other large-scale, agent-based phenomena such as epidemic spread, market dynamics, and urban systems where individual heterogeneity matters at scale. Future work will explore richer agent memory and lifelong learning, multi-modal interactions, and tighter integration with real-world data streams, with the longer-term goal of developing population-scale digital twins that complement, rather than replace, empirical inquiry.

\section{Methods}
\label{sec:methods}

\subsection{Framework formalism}
\label{sec:framework_formalism}

\simulator is designed as a modular framework that integrates LLM-powered social cognition, modeling social dynamics as transitions between agent and environment states orchestrated by simulation operations and scheduled through an event queue. The simulation model is defined as the tuple
\begin{align}
M &:= \langle D, T, S_{A}, S_{E}, V, Q, F \rangle \\
F &:= \{ f_{I}, f_{P}, f_{\Pi}, f_{A}, f_{E}, f_{U} \}.
\end{align}
Here, $T$ denotes the timeline of the simulation. At each time step, the system maintains a dynamic set of active agents, which may change over time due to internal events such as birth, death, or shifts in participation (for example, when individuals join or leave a social platform). Each agent has a state $S_A$ composed of three elements: a static profile that remains constant throughout the simulation (such as demographics or personality traits), an internal status that evolves over time (capturing aspects like memory, beliefs, and goals), and an external status that reflects observable features like the agent's location or social connections. The environment state $S_E$ is defined by both static components, such as spatial layout or physical rules, and dynamic components that change as the simulation progresses, such as weather conditions or disasters.

Events $v \in V$ in \simulator serve as the primary drivers of simulation dynamics. Each event encapsulates a discrete interaction or change in the system, such as an agent sending a message, relocating, or forming a relationship. Events are assigned timestamps for execution and may involve one or more initiating entities and one or more target entities. They carry structured payloads that describe the content and context of the interaction, for instance the text of a message, a resource transfer, or a behavioral intention. Events may also include a priority level, allowing the system to resolve multiple simultaneous events deterministically. All events are stored in the event queue $Q$, implemented as a min-heap over $(\text{time}, \text{priority}, \text{sequence})$ tuples, ensuring consistent and temporally coherent execution of interactions throughout the simulation.

The core dynamics of \simulator are governed by a collection of simulation operations $F$ powered by large language models. The initialization operation $f_I$ configures the simulation by generating the initial agent states, the environment state, and a set of starting events from a seed dataset $D$, with various initialization methods including rule-based generation~\citep{chopra2024limits}, LLM-based generation~\citep{xie2024can}, and data augmentation~\citep{yang2024oasis}. During simulation, agents perceive their surroundings via a perception operation $f_P$, and make decisions and generate actions using a policy operation $f_\Pi$ that produces new events. As time progresses, agent and environment states evolve independently between events through dedicated evolution operations $f_A$ and $f_E$, reflecting internal dynamics such as memory decay or environmental drift. An update operation $f_U$ applies events back to the simulation system, ensuring consistent state transitions. After the simulation, a readout process aggregates historical system states and events into measurable outputs for downstream analysis.

\subsection{Optimization implementation}
\label{sec:optimization_impl}

To support efficient inference at planetary scale, \simulator implements a multi-tiered optimization pipeline. At the inference layer, prompt caching reduces redundant LLM invocations by reusing responses for repeated or near-identical queries, supported by both an exact (SHA-256-keyed LRU) cache and a FAISS-based semantic cache. Knowledge distillation is applied to train compact, task-specific surrogate models that approximate LLM behavior for routine decisions and environment updates, offloading frequent calls from full-scale LLMs while preserving behavioral fidelity. Simulation operations are dispatched through a mixture-of-models architecture, in which a configurable routing policy assigns each request to a backend from a registered table that may include full-scale LLMs accessed through OpenAI-compatible APIs, distilled surrogate models, and precomputed lookup tables. \simulator currently supports three routing policies, namely all-LLM, all-surrogate, and per-sample weighted routing. Backends accessed through this layer expose an asynchronous interface that supports vectorized batch dispatch, allowing the same router to drive both per-prompt LLM invocations and batched surrogate inference.

At the runtime layer, \simulator leverages compressed graph structures and vectorized batch operations to support efficient state updates and neighbor queries across billion-node networks. Adjacency information is stored in compressed sparse row (CSR) format and serialized to HDF5 for efficient access at billion-node scale. Agent state can be held in a per-agent record store for small-scale experiments or in a columnar store backed by per-field numpy arrays (optionally HDF5-mapped) for billion-agent runs, where vectorized per-field updates dominate. The simulation logic is driven by an event queue implemented as a min-heap over $(\text{time}, \text{priority}, \text{sequence})$ tuples. Same-(time, priority) events are popped together as a concurrent batch, grouped by event kind, and dispatched to the relevant policy or update operation. The resulting modular and composable architecture supports plug-and-play accelerations, allowing domain-specific enhancements to be integrated without altering the core simulation logic.

\subsection{WVS-based agent profile construction}
\label{sec:wvs_profile_construction}

Most agent profiles used in this work were derived from the publicly available World Values Survey (WVS) Wave 7 dataset (2017--2022)~\citep{haerpfer2022world}. The WVS offers broad demographic and attitudinal data (Supplementary Fig.~\ref{fig:supp_trust}d), providing a globally representative empirical foundation for constructing realistic agent personas. Following the methodology of WorldValuesBench~\citep{zhao2024worldvaluesbenchlargescalebenchmarkdataset}, we extracted a wide range of socio-demographic attributes, including geographic location, gender, age, migration status, education, employment, income, religion, ethnicity, and subjective social class. Records with incomplete responses on core fields (e.g., gender, age) were excluded, yielding a cleaned dataset of 96{,}125 valid entries. Each cleaned WVS record was transformed into a natural-language agent profile written in the second person (``You are\dots'') to facilitate character embodiment by the LLMs and to include rich socio-demographic context; an example profile is given in Appendix~\ref{app:wvs_profile_example}.

\subsection{Trust Games experimental setup}
\label{sec:trust_games_setup}

Three Trust Game variants were implemented, all simulated with \texttt{gemini-2.0-flash-001} unless stated otherwise: the canonical one-shot Trust Game, in which a trustor sends an integer amount $N \in [0, 10]$ from a \$10 endowment to an anonymous trustee who receives $3N$ and decides how much to return; the Dictator Game, in which the trustee has no return option and the trustor's transfer is purely altruistic; and a Repeated Trust Game, in which the same pair of agents play seven rounds of the canonical setup with each round restarting the trustor's endowment at \$10. Cross-model comparisons in Fig.~\ref{fig:fig2_trust}c--d additionally use \texttt{gpt-4.1-nano} as a contrast.

Prompts were designed to (i) define the agent's role (trustor or trustee), (ii) specify the game rules, and (iii) constrain output to a strict JSON object with a brief reasoning string and an integer amount. The full prompt templates for the canonical Trust Game (trustor and trustee), the Dictator Game (trustor), and the Repeated Trust Game (trustor and trustee) are reproduced in Appendix~\ref{app:trust_prompt_templates}.

\subsection{Opinion diffusion experimental setup}
\label{sec:opinion_diffusion_setup}

We generate a Barab\'{a}si--Albert (BA) scale-free network with one billion nodes using the igraph library (\texttt{ig.Graph.Barabasi} with \texttt{implementation="psumtree"}). Each new node attaches to $m = 3$ existing nodes via preferential attachment, producing a power-law degree distribution $P(k) \sim k^{-3}$. The resulting adjacency structure is stored as a compressed sparse row (CSR) matrix and converted to HDF5 format for efficient random access during simulation. Nodes are ranked by degree: the top 20\% (200~million) serve as influencers, and the remaining 80\% (800~million) as influencees. Inter-influencer edges are filtered out so that each influencer's neighbor list contains only influencee nodes, ensuring a unidirectional influence flow.

Each agent is assigned a demographic profile sampled from a fixed pool of 10{,}000 profiles drawn from the cleaned WVS dataset described in Section~\ref{sec:wvs_profile_construction}. Each agent is given a controversial statement and assigned an initial opinion in $\{$\texttt{agree}, \texttt{disagree}, \texttt{neutral}$\}$. Influencee opinions are initialized uniformly at random; influencer opinions are configured under three seeding schemes: \textbf{1D1N} (50\% \texttt{disagree}, 50\% \texttt{neutral}), \textbf{1A1N} (50\% \texttt{agree}, 50\% \texttt{neutral}), and \textbf{Random}.

Direct LLM inference at billion-agent scale is infeasible, so the policy operation is resolved by a knowledge-distilled surrogate (the embedding-MLP architecture described in Section~\ref{sec:surrogate_training}) wrapped in a precomputed lookup table. Each of the 10{,}000 demographic profiles is pre-embedded once using OpenAI \texttt{text-embedding-3-large} ($d = 3072$); the surrogate takes the concatenation of the influencer embedding, target embedding, and one-hot encodings of the two stances (yielding a 6{,}150-dimensional input) and outputs a 3-class softmax over $\{$\texttt{disagree}, \texttt{neutral}, \texttt{agree}$\}$. To eliminate inference overhead during the billion-agent simulation, we precompute a four-dimensional lookup table $T[i_{\text{prof}}, t_{\text{prof}}, i_{\text{stance}}, t_{\text{stance}}] \to \text{final\_stance}$ of shape $[10{,}000 \times 10{,}000 \times 3 \times 3]$, covering all 900~million possible input combinations; during simulation, each interaction reduces to a single array lookup.

Driven by \simulator's asynchronous event queue, the simulation proceeds for 100 rounds. In each round, 1\% of all influencers (approximately 2~million nodes) are randomly sampled and each emits an influence event targeting every one of its influencee neighbors, all sharing the same timestamp. \simulator's event-aggregation mechanism merges these concurrent events into a single batch, which the policy operation resolves through one vectorized lookup on the prediction table using both agents' profile indices and current stances. Affected influencee stances are updated when the batch returns, and per-round opinion distributions and stance transition counts are recorded for downstream analysis.

\subsection{Surrogate training and selection}
\label{sec:surrogate_training}

For the sensitivity analysis in Section~\ref{sec:analysis}, we trained five surrogate architectures on the same teacher-LLM interaction samples per topic. Two architectures operate on raw WVS demographic attributes: a softmax regression (SR) and a gradient-boosted tree ensemble (LightGBM~\citep{ke2017lightgbm}). Two operate on pre-computed \texttt{text-embedding-3-large} profile embeddings: a multilayer perceptron (MLP) with hidden sizes $(512, 256)$, ReLU activations, and dropout 0.1, and a Transformer encoder~\citep{vaswani2017attention} over the embedding-stance feature sequence. The fifth architecture is a Qwen3-0.6B small language model~\citep{qwen3technicalreport} (SLM) supervised-fine-tuned on the same teacher-LLM interaction samples using the same prompt template as the teacher LLM, so that it consumes natural-language input and produces a generative stance label.

The teacher LLM is Gemini~2.0~Flash. For each topic, approximately 400{,}000 interaction tuples (influencer profile, target profile, influencer stance, target stance) are drawn from the cleaned WVS pool and resolved through the teacher LLM under the same prompt used in the billion-agent simulation. The four classifier surrogates are trained with cross-entropy loss; the SLM is fine-tuned with the standard next-token-prediction loss on the teacher's interaction transcripts.

As reported in Section~\ref{sec:analysis}, per-sample classification accuracy alone is not a reliable indicator of surrogate quality, because checkpoints with similar F1 can differ several-fold in their distributional gap from the teacher LLM. Selection therefore combines two criteria: per-sample macro-F1 on a held-out interaction set, and the change-rate gap, defined as the absolute difference between the surrogate's fraction of stance-changing interactions and the teacher LLM's fraction, evaluated over the same held-out set. We first shortlist the top-$N$ checkpoints by macro-F1 and then choose, within this shortlist, the checkpoint whose change-rate gap is smallest. The embedding-MLP checkpoint selected by this procedure is the surrogate used in the billion-agent simulation reported in Section~\ref{sec:1b}.

\subsection{Network topology experimental setup}
\label{sec:network_topology_setup}

All simulations in the network-topology experiment are run at a fixed scale of $N = 1{,}000$ agents across six network substrates. Three sparse synthetic graphs serve as conventional baselines: a Barab\'{a}si--Albert graph with attachment parameter $m = 3$, an Erd\H{o}s--R\'{e}nyi graph with target mean degree $\langle k \rangle = 6$, and a uniformly random spanning tree. Two density-matched synthetic graphs increase their generation parameters so that the total number of edges matches the empirical substrate: a Barab\'{a}si--Albert graph with $m = 27$ and an Erd\H{o}s--R\'{e}nyi graph with $\langle k \rangle = 54$. The empirical substrate is a connected 1{,}000-node subgraph sampled from the Twitch-DE user-friendship graph released in the MUSAE dataset~\citep{rozemberczki2021multi}, which contains 9{,}498 nodes and 153{,}138 undirected edges in full. To obtain the 1{,}000-node sample, the highest-degree node of the original graph is used as the seed of a degree-prioritized breadth-first search; at each expansion step the currently uncovered neighbors with the largest degree are visited first, and the search halts once 1{,}000 nodes have been reached. The induced subgraph has $\langle k \rangle = 54.3$, clustering coefficient $C = 0.31$, and $E = 27{,}152$ edges.

Each agent maintains two continuous internal states: a stance $s \in [-1, +1]$, where negative values indicate skepticism and positive values support, and a confidence $c \in [0, 1]$ that modulates how willingly the agent updates in response to incoming messages. Agents classified as supporters, skeptics, and neutrals correspond to $s > 0.2$, $s < -0.2$, and $-0.2 \le s \le 0.2$, respectively. At initialization, stances are sampled uniformly at random in $[-1, +1]$ and confidences are set to a moderate baseline.

Driven by \simulator's asynchronous event queue, each simulation runs for 20 days with 10 event slots per day. Each slot dispatches two speaker-activation events under an expression-desire heuristic; each activated speaker fans out heard events to its graph neighbors, all sharing the same timestamp, which \simulator's event-aggregation mechanism merges into a single concurrent batch. Each heard event triggers an LLM-powered state update on the listener, producing a new stance and confidence conditional on the received message. At the close of each day, a reflective-update event is enqueued for every agent that received at least one heard event that day; the resulting same-timestamp batch prompts the LLM to integrate the day's accumulated exposures and produce an updated internal state per agent. All LLM-powered updates are executed by \texttt{deepseek-v4-flash}.

\subsection{Related work}
\label{sec:related-work}

Recent advances in large language models (LLMs) have significantly reshaped agent-based social simulations, enabling more realistic representations of human cognitive processes and social behaviors~\citep{openai2024gpt4technicalreport,Wang_2024}. Leveraging extensive world knowledge and sophisticated natural language capabilities, LLM-driven agents can simulate complex human interactions, including context-aware decision-making and adaptive social responses~\citep{aher2023using}. These virtual societies, powered by LLM agents, have successfully exhibited emergent phenomena such as collective information diffusion and realistic economic market dynamics~\citep{li2023econagent,park2023generative}.

Agent-based modeling (ABM) has historically served as a foundational approach for exploring complex societal phenomena through individual-level interactions. Early influential studies demonstrated how simple behavioral rules could generate complex macro-level outcomes, such as residential segregation arising from mild individual preferences~\citep{schelling1969models} and cultural diversity emerging from local interactions~\citep{axelrod1997dissemination}. Later, more sophisticated ABMs like Sugarscape~\citep{epstein1999agent} expanded agent capabilities, and Palmer et al.'s artificial stock market~\citep{palmer1999artificial} employed adaptive agents to simulate realistic market dynamics. Despite their utility, traditional ABMs face inherent limitations including oversimplified agent behaviors constrained by manually crafted rules~\citep{colledanchise2019learning}, subjective parameter selection, and scalability challenges when modeling large populations~\citep{macal2018chisim}. Reinforcement learning approaches~\citep{ie2019recsim,arulkumaran2019alphastar,mnih2013playing} address some of these limitations by automating agent decision-making, but face significant challenges when extended to open-ended social simulations, primarily due to difficulties in specifying appropriate reward functions that accurately reflect complex human motivations~\citep{chen2020knowledge,lv2021user,wang2023multi}.

Beyond demonstrating emergent social phenomena, subsequent research has expanded the applications of LLM-based agents across diverse domains. Studies have explored opinion dynamics~\citep{chuang2024simulating,liu2023training}, misinformation propagation~\citep{acerbi2023large,gao2023s3}, geopolitical conflicts~\citep{hua2024waragent}, and political election forecasting~\citep{zhang2024electionsim}. LLM agents can also replicate classical experiments from economics, psychology, and social sciences, capturing intricate human-like decision patterns and biases~\citep{aher2023using}. Multimodal extensions process and respond to visual, auditory, and textual stimuli, enabling richer simulations of human behavior across diverse contexts such as urban environments, healthcare, and economic decision-making~\citep{li2025agent,liu2024lmagent,xu2023urban}.

Despite these advances, scaling LLM-based simulations to large populations remains computationally challenging. Most existing frameworks are limited to relatively small groups of agents (often fewer than 1{,}000) due to significant computational overhead~\citep{tang2024gensimgeneralsocialsimulation,wang2024user}. Recent efforts have begun addressing these scalability challenges, developing architectures capable of efficiently simulating large-scale agent societies using parallel computation and modular designs~\citep{liu2024lmagent,piao2025agentsociety,yang2024oasis,mou2024individual,wang2025yulanonesimgenerationsocialsimulator}. Significant challenges persist, however, particularly in efficiently orchestrating large-scale LLM inference and accurately representing complex societal dynamics at planetary scales~\citep{zhang2025socioverse}. \simulator contributes directly to this area through joint algorithmic and system optimizations, particularly a mixture-of-models inference engine that combines full LLMs with distilled surrogates, enabling efficient simulation at billion-agent scale.

\section{Acknowledgements}
\label{sec:acknowledgements}

We thank Zhongguancun Academy and Zhongguancun Institute of Artificial Intelligence for providing and maintaining the computational resources that made this work possible.

\section{Author contributions}
\label{sec:author_contributions}

H.G., J.H., X.Y., Y.C., X.X., S.Z., Y.G., E.C., T.-Y.L. and Z.L. conceived and supervised the study. H.G., J.H. and L.F. designed the framework and methodology. H.G., J.H., L.F., Z.R. and S.H. implemented the software, conducted the experiments, and performed the analyses. H.G., J.H., L.F., Z.R., S.H., X.Y., Y.C., S.Z., Y.G. and Z.L. wrote and revised the manuscript. All authors reviewed and approved the final version of the manuscript.

\clearpage

\clearpage

\clearpage

\clearpage
\bibliography{sn-bibliography}%

\clearpage
\begin{appendices}

\section{Supplementary Figures}

\begin{figure}[h!]
    \centering
    \includegraphics[width=0.99\textwidth]{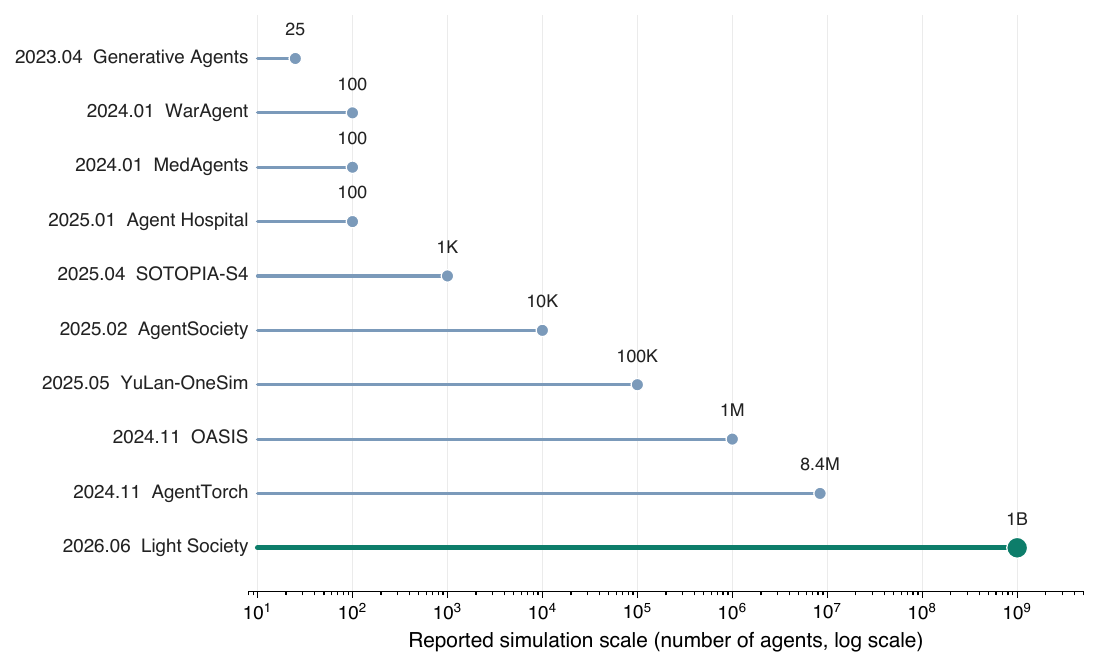}
    \caption{\textbf{Reported simulation scales across representative works.} Data are collected from the experimental scales reported in the corresponding papers. Prior systems in our comparison range from tens of agents to roughly $10^7$ agents, whereas \simulator reaches $10^9$ agents.}
    \label{fig:supp_scale_comparison}
\end{figure}

\begin{figure}[h!]
    \centering
    \includegraphics[width=0.99\textwidth]{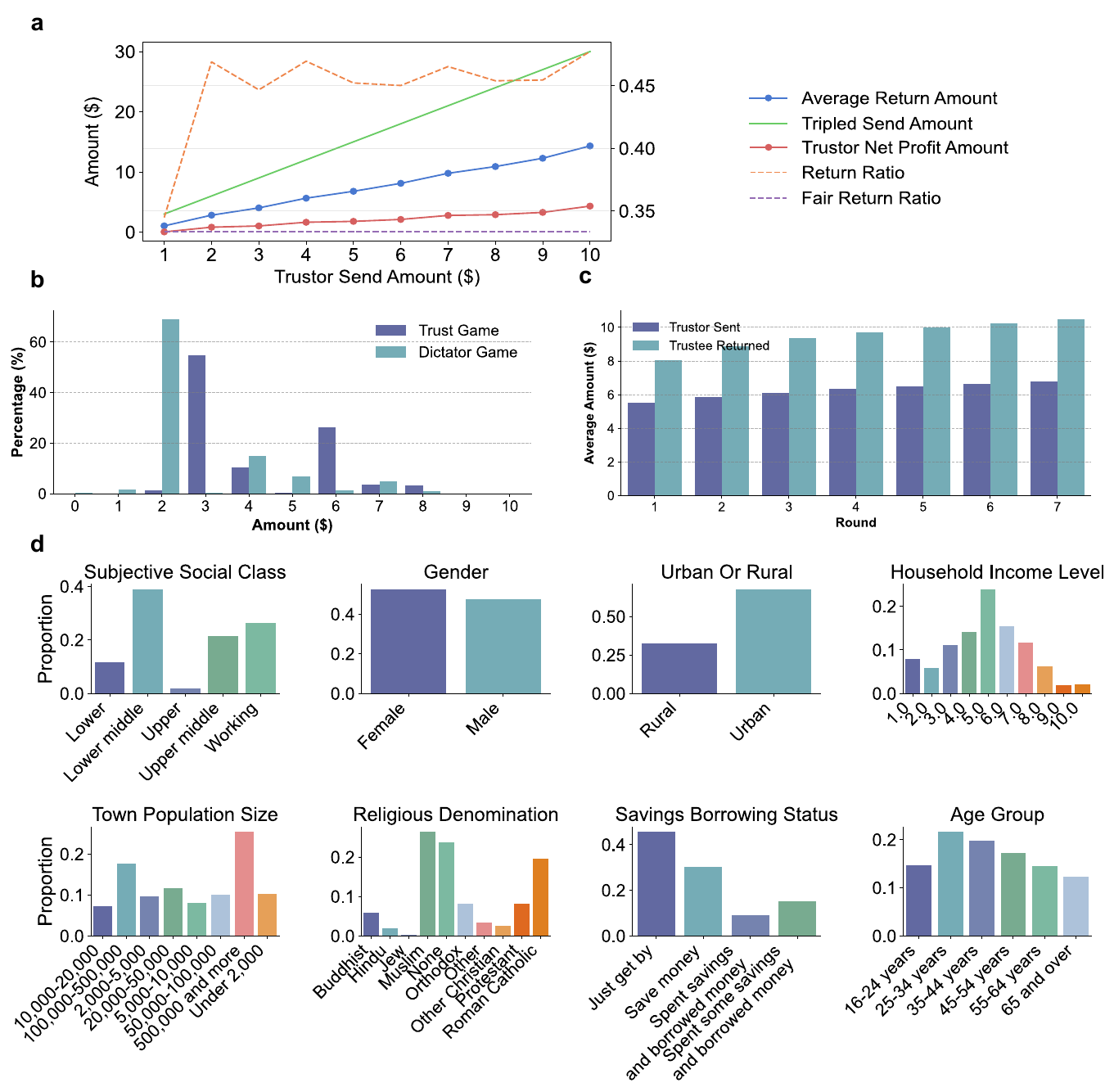}
    \caption{\textbf{Supplementary results for Trust Games.}
\textbf{a}, Trustee return behavior across all received amounts (tripled from trustor sending), showing average return, trustor net profit, fair return line, and return ratios.
\textbf{b}, Distribution of trustor sending amounts in the canonical Trust Game versus the Dictator Game. Trustors tend to send higher amounts when reciprocity is possible (Trust Game) compared to purely altruistic transfers (Dictator Game), highlighting the role of reciprocal expectations in promoting prosocial behavior.
\textbf{c}, Average trustor sending and trustee returning amounts across seven rounds of the Repeated Trust Game. Both amounts increase over repeated interactions, indicating that trust and reciprocity strengthen as agents accumulate interaction history and develop stable cooperative norms.
\textbf{d}, Distribution of eight key socio-demographic variables in the constructed WVS-based agent dataset used for simulation.}
    \label{fig:supp_trust}
\end{figure}

\begin{figure}[h!]
    \centering
    \includegraphics[width=0.99\textwidth]{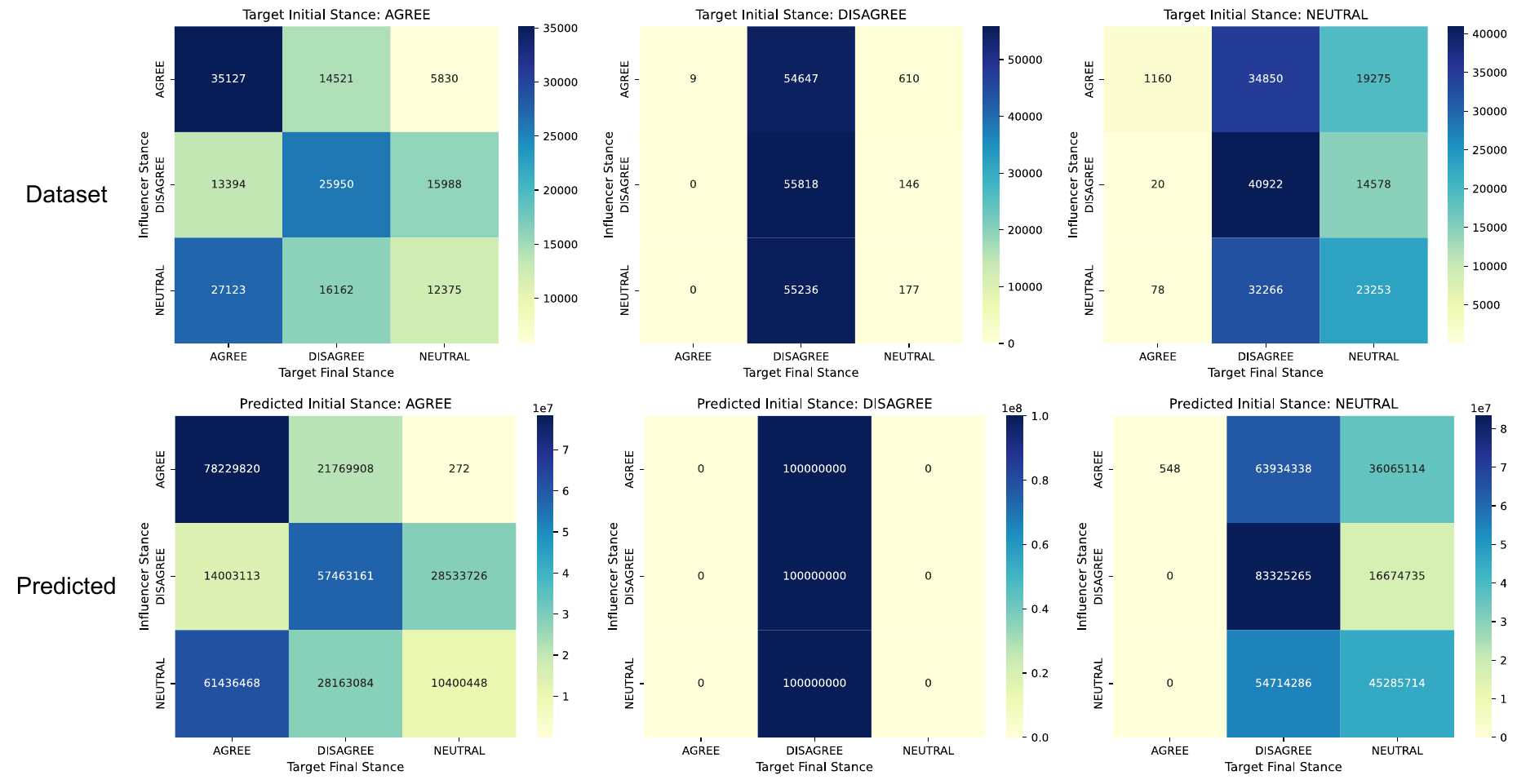}
    \caption{\textbf{Stance transition distributions for the topic ``The Earth is flat.''}
\textbf{Top row}, Ground-truth transition counts from LLM-generated interaction samples, conditioned on target initial stance (Agree, Disagree, Neutral). Each heatmap shows influencer stance (rows) versus target final stance (columns).
\textbf{Bottom row}, Corresponding transition counts predicted by the surrogate model over all profile pairs.}
    \label{fig:supp_trans_earth}
\end{figure}

\begin{figure}[h!]
    \centering
    \includegraphics[width=0.99\textwidth]{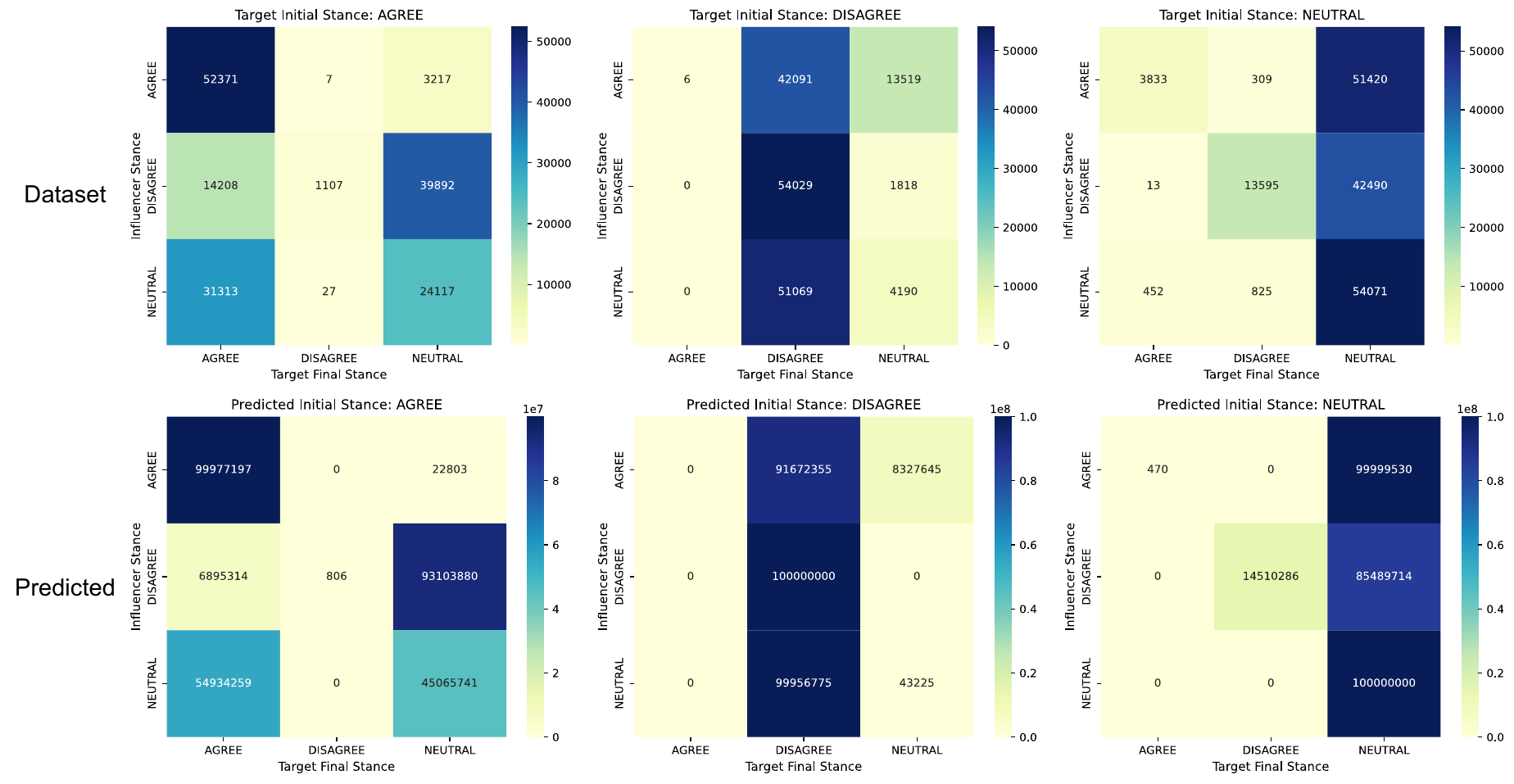}
    \caption{\textbf{Stance transition distributions for the topic ``Humans will establish a Martian city within 50 years.''}
\textbf{Top row}, Ground-truth transition counts from LLM-generated interaction samples.
\textbf{Bottom row}, Surrogate model predictions.}
    \label{fig:supp_trans_martian}
\end{figure}

\begin{figure}[h!]
    \centering
    \includegraphics[width=0.99\textwidth]{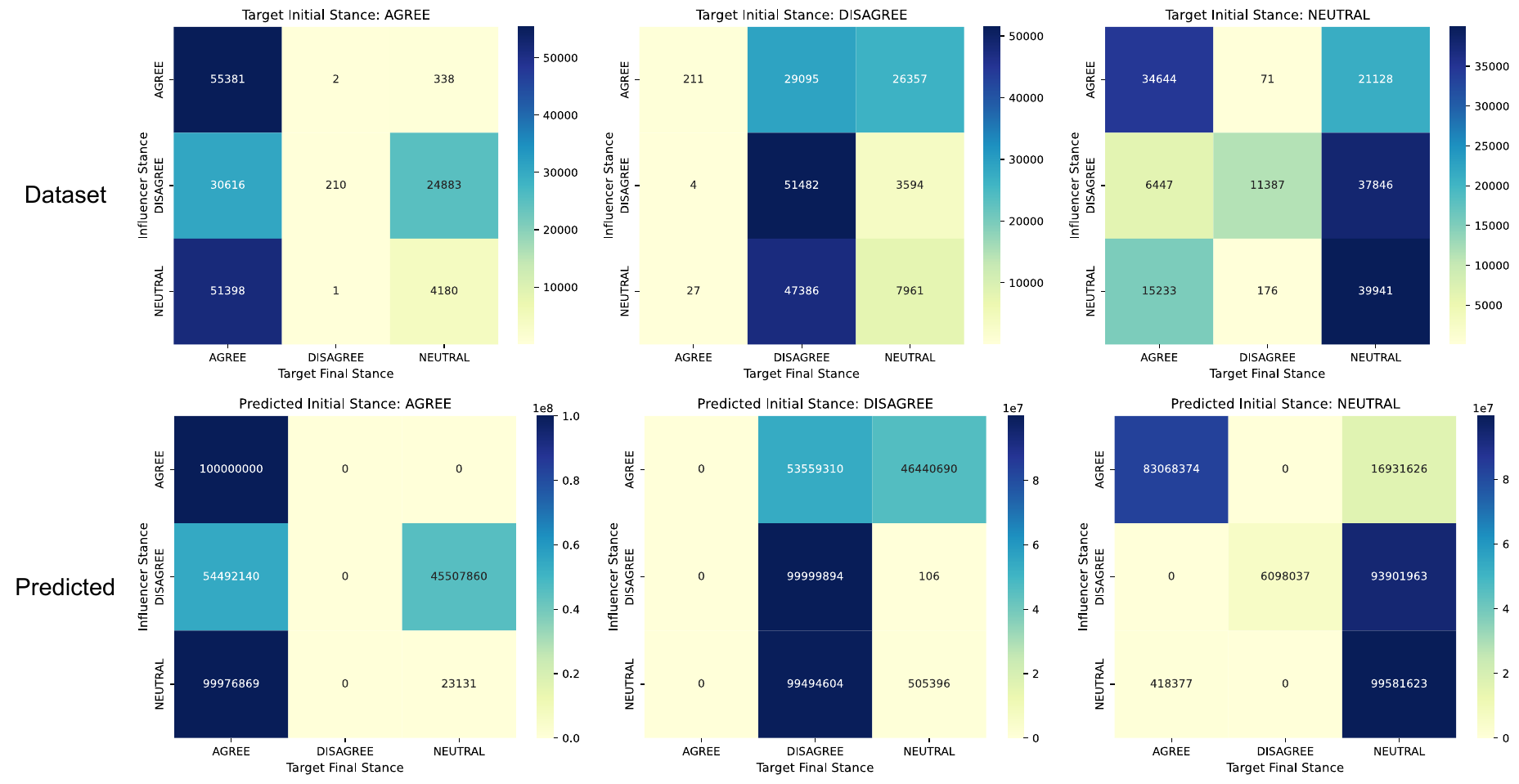}
    \caption{\textbf{Stance transition distributions for the topic ``Short-form videos are reducing human attention spans.''}
\textbf{Top row}, Ground-truth transition counts from LLM-generated interaction samples.
\textbf{Bottom row}, Surrogate model predictions.}
    \label{fig:supp_trans_video}
\end{figure}

\begin{figure}[h!]
    \centering
    \includegraphics[width=0.99\textwidth]{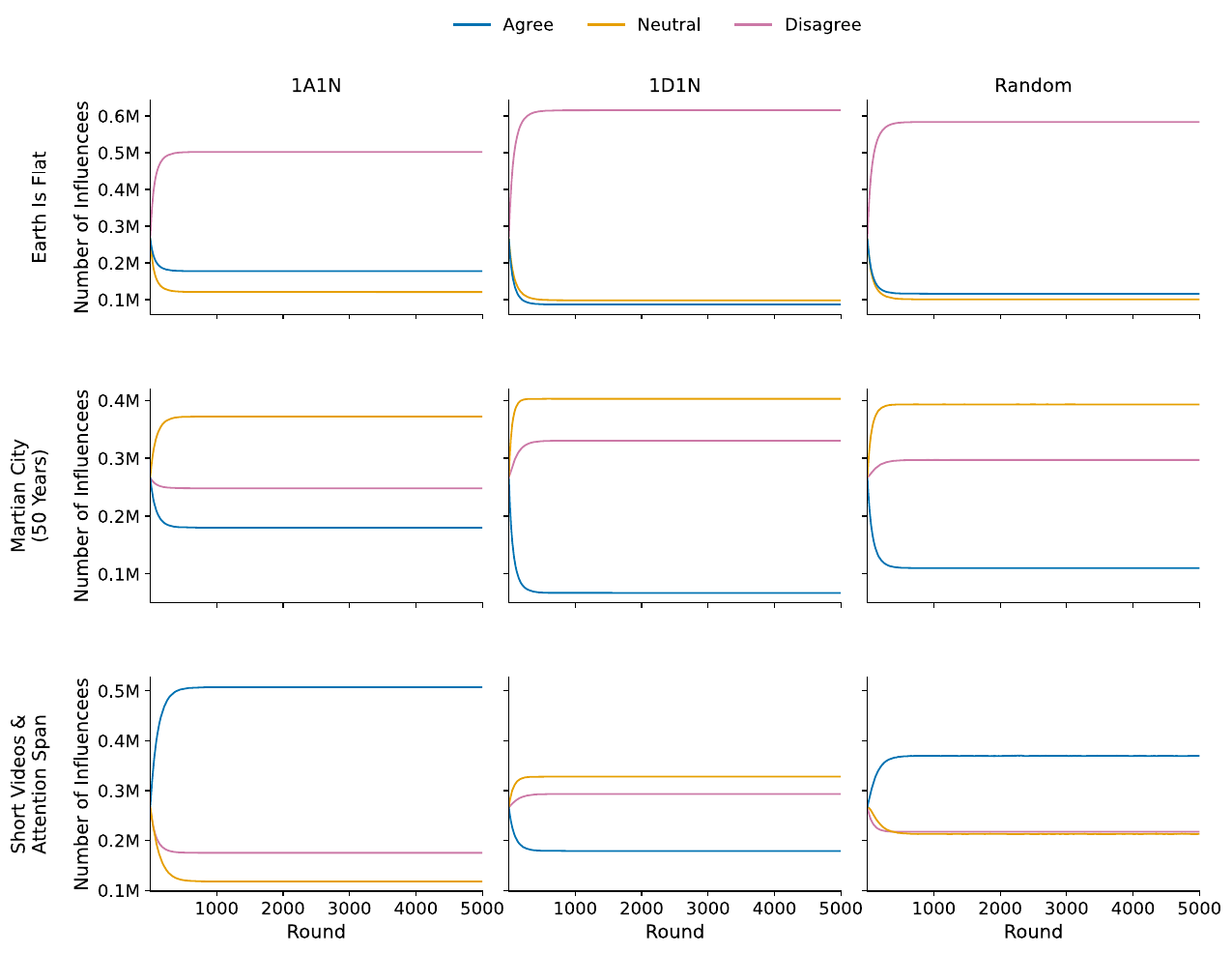}
    \caption{\textbf{Long-horizon opinion diffusion over 5{,}000 simulation rounds.} Rows correspond to the three topics (top: Flat-Earth theory; middle: Martian city; bottom: short-form videos), and columns correspond to the three seeding schemes (left: 1A1N; middle: 1D1N; right: Random). Across all nine topic-scheme combinations, the opinion distributions eventually settle into a dynamic equilibrium.}
    \label{fig:supp_long_horizon}
\end{figure}

\begin{figure}[h!]
    \centering
    \includegraphics[width=0.99\textwidth]{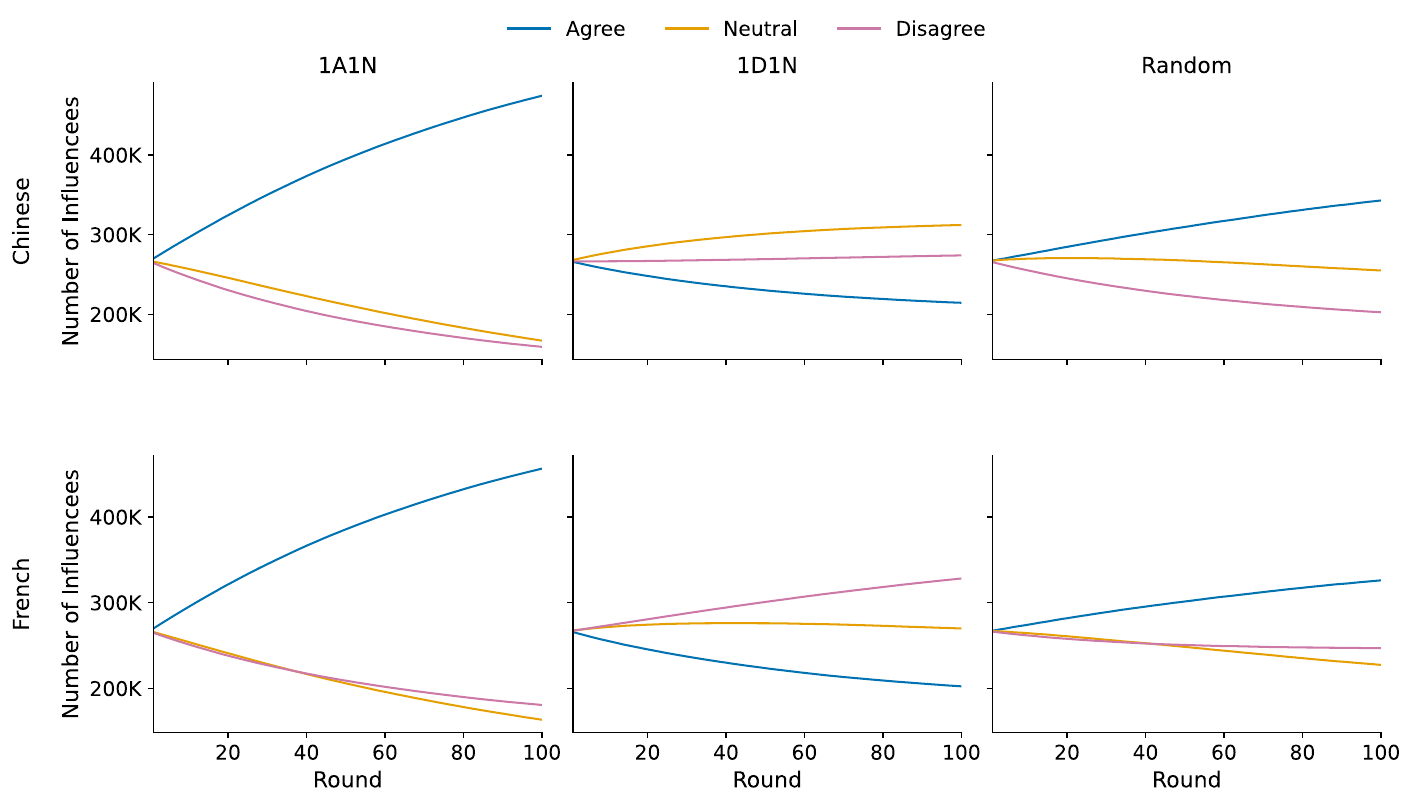}
    \caption{\textbf{Effect of communication language on opinion diffusion for the topic ``Short-form videos are reducing human attention spans.''} Rows correspond to the communication language used by the agents (top: Chinese; bottom: French), and columns correspond to the three seeding schemes (left: 1A1N; middle: 1D1N; right: Random). Within each seeding scheme, the resulting opinion trajectories differ between the two languages, indicating that the choice of communication language measurably influences the simulated opinion dynamics.}
    \label{fig:supp_lang_diff}
\end{figure}

\begin{figure}[h!]
    \centering
    \includegraphics[width=0.99\textwidth]{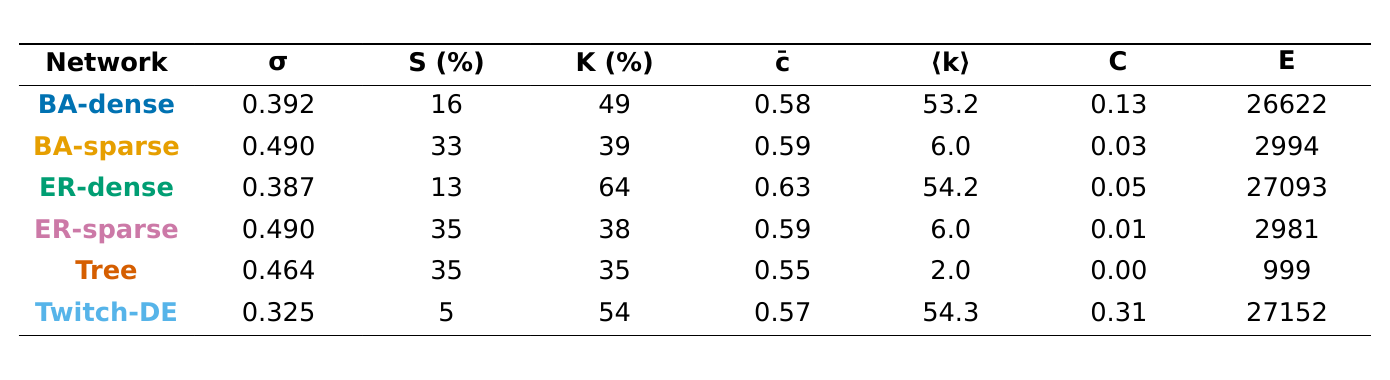}
    \caption{\textbf{Per-topology summary statistics for the network topology experiment.} Rows correspond to the six network substrates used in Section~\ref{sec:effect_of_social_network_topology}, three sparse synthetic graphs, two density-matched dense synthetic graphs, and the empirical Twitch-DE subgraph. Dynamic outcomes at day 20 are reported in the first four columns: $\sigma$, standard deviation of the per-agent stance; $S$, fraction of supporters ($s > 0.2$); $K$, fraction of skeptics ($s < -0.2$); and $\bar c$, mean agent confidence. The last three columns summarize the structural properties of each graph: mean degree $\langle k \rangle$, clustering coefficient $C$, and number of edges $E$.}
    \label{fig:supp_network_table}
\end{figure}

\begin{figure}[h!]
    \centering
    \includegraphics[width=0.7\textwidth]{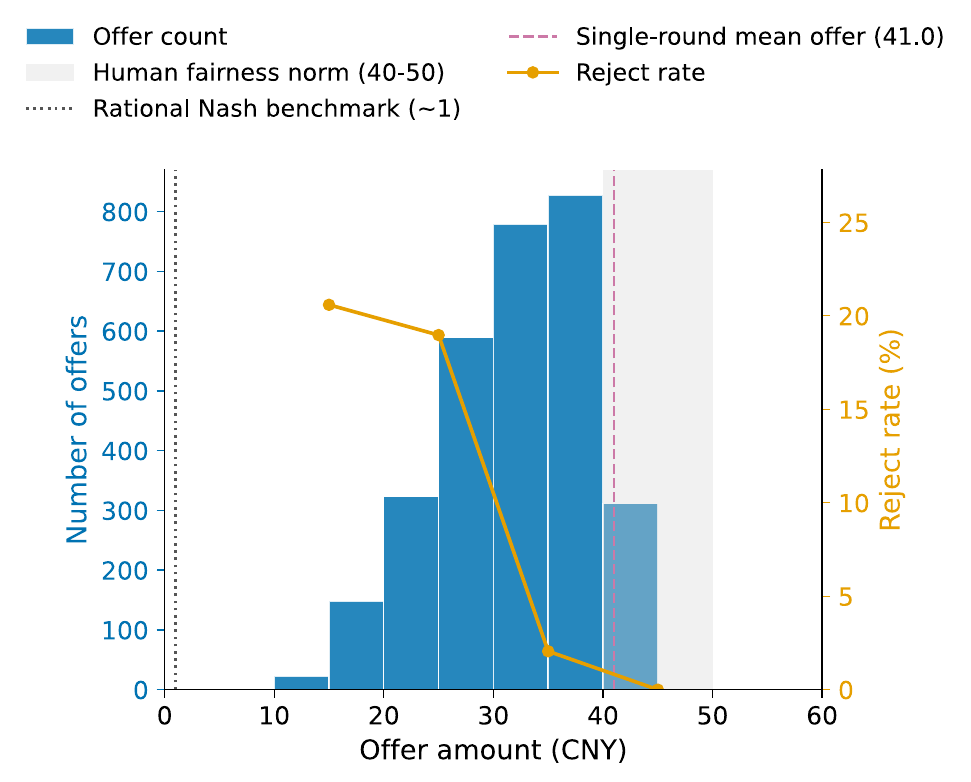}
    \caption{\textbf{Aggregate offer distribution and offer-conditioned rejection risk in the 10-round Ultimatum Game.} Blue bars: histogram of all 3{,}000 offers across 300 proposer-responder pairs and ten rounds. Orange curve: rejection rate within each 10-CNY offer bin. Dotted vertical line: game-theoretic Nash benchmark (offer of 1 out of 100). Dashed vertical line: single-round reference mean offer (41.0). Shaded band: the 40\%--50\% fair-offer range commonly observed in human Ultimatum-Game experiments.}
    \label{fig:supp_ultimatum}
\end{figure}

\begin{figure}[h!]
    \centering
    \includegraphics[width=0.99\textwidth]{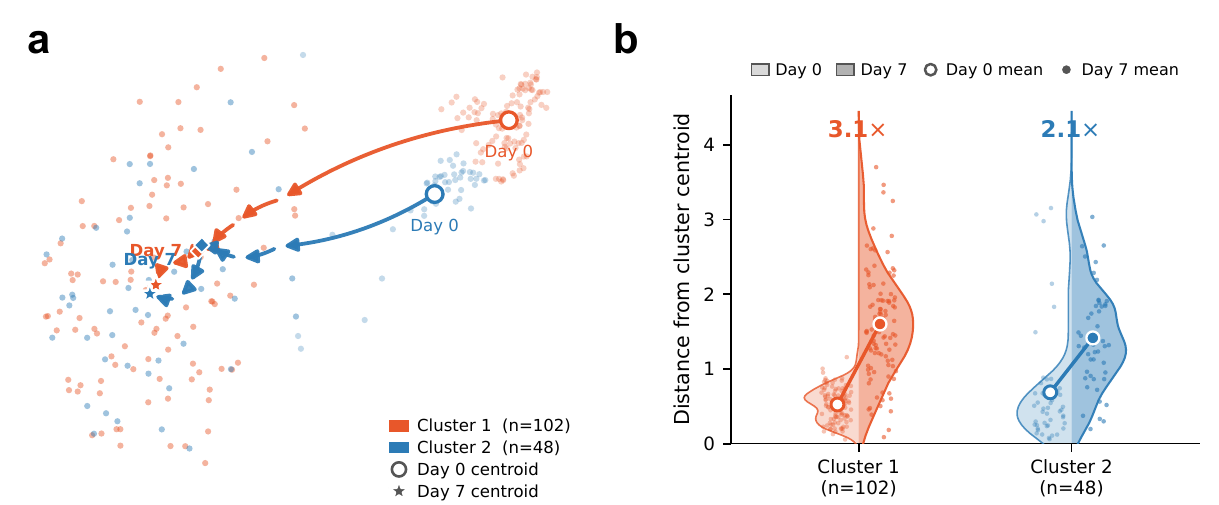}
    \caption{\textbf{Memory-augmented free-discussion simulation on the ``Mars city in 50 years'' topic.} 150 agents, seven simulation days. \textbf{a}, Centroid trajectories of the two initial opinion clusters in a 2D UMAP projection of the per-agent embedding; open circles mark Day 0, stars mark Day 7. The two centroids move toward a shared region in semantic space. \textbf{b}, Within-cluster spread, measured as the distance from each cluster's centroid, at Day 0 (left half-violin) and Day 7 (right half-violin) for each cluster, with the Day-0-to-Day-7 mean ratio annotated above.}
    \label{fig:supp_free_discuss}
\end{figure}

\clearpage
\section{Supplementary Notes}

\subsection{Example WVS agent profile}
\label{app:wvs_profile_example}

The following example shows a representative agent profile generated from the cleaned WVS Wave 7 dataset, used as the input persona for Trust Games and the other experiments described in the main text.

\begin{tcolorbox}[colback=blue!5!white, colframe=blue!75!black, breakable,title=An Example from WVS Dataset]
You are a Female, 60 years old person. You live in AD: Andorra la Vella, Andorra. Your town has a population of 20,000-50,000, is classified as a Capital city, is considered Urban. You are an immigrant to this country (born outside this country) born in Spain. Your citizenship status is ``Not, I am not a citizen of this country''. Your native language is Spanish; Castilian. Your ethnicity/race is AD: Caucasian white. You don't have any religious beliefs. You are Married and have 2 children. Your highest education level is Upper secondary education (ISCED 3), your father's education level is Upper secondary education (ISCED 3), your mother's education level is Upper secondary education (ISCED 3), your spouse's education level is Upper secondary education (ISCED 3). You are Full time (30 hours a week or more) working as Sales (for example: sales manager, shop owner, shop assistant, insurance agent, buyer) in the Private business or industry sector. Your spouse is Full time (30 hours a week or more). Your father's occupation was Semi-skilled worker (for example: bricklayer, bus driver, cannery worker, carpenter, sheet metal worker, baker), your mother's occupation was Farm worker (for example: farm labourer, tractor driver). Your father is Immigrant born in Spain. Your mother is Immigrant born in Spain. You are not the main income earner in your household. Your financial situation: Spent some savings and borrowed money. You consider yourself to be Lower middle class. On a scale of 1-10, You place your household income at level 5.
\end{tcolorbox}

\subsection{Supplementary results for Trust Game variants}
\label{app:trust_variants_results}

Supplementary Fig.~\ref{fig:supp_trust}b compares trustor sending amounts between the canonical Trust Game and the Dictator Game. When trustees have the option to reciprocate, trustors send notably higher amounts, suggesting that expectations of reciprocity and mutual benefit encourage greater initial trust. In contrast, the Dictator Game elicits lower transfers because giving is motivated purely by altruism rather than strategic trust.

Supplementary Fig.~\ref{fig:supp_trust}c shows average sending and returning amounts across seven repeated interactions of the Repeated Trust Game. Both values increase steadily over time, reflecting the emergence of cooperative norms and the gradual buildup of interpersonal trust. Such dynamics are consistent with findings from human behavioral experiments~\citep{cochard2004trusting}, where repeated interactions reduce uncertainty, reward reciprocal behavior, and facilitate stable cooperative relationships. Together, these supplementary results show that \simulator captures context-dependent social behaviors and reproduces how repeated interactions and reciprocal incentives shape trust and cooperation.

\subsection{Trust Game prompt templates}
\label{app:trust_prompt_templates}

The following prompt templates were used for the canonical Trust Game (trustor and trustee), the Dictator Game (trustor), and the Repeated Trust Game (trustor and trustee). Each template defines the agent's role and game rules and constrains the model output to a strict JSON object with a brief reasoning string and an integer amount.

\begin{tcolorbox}[colback=blue!5!white, colframe=blue!75!black, breakable,title=Trustor Prompt (Trust Game)]
You are participating in a one-shot, anonymous economic game. Here is your profile: \{profile\_text\}\\

Experiment Instructions for you:

1. You begin this interaction with \$10.

2. You need to decide how much money, N, to send to another anonymous participant. N must be an integer between \$0 and \$10, inclusive.

3. The amount N you choose to send will be tripled, and the other participant will receive 3N.

4. After receiving the tripled amount, the other participant will decide how much of that money (3N) to send back to you.

5. This is a single interaction; you will not interact with this participant again.\\

Your Decision:

Based on your profile and the instructions, decide the amount N (from 0 to 10) you want to send.\\

Output Format:

Provide your decision *only* as a JSON object with two keys:

- ``thinking\_process'': A brief explanation (1-2 sentences) of your reasoning for choosing the amount N, considering your profile and the situation.

- ``amount'': The integer amount N (0-10) you decided to send.\\

Example Output:

\{\{

``thinking\_process'': ``Given my situation and the anonymous nature, I feel a bit cautious, but I'll send a moderate amount hoping for a fair return.'',

``amount'': 5

\}\}\\

Now, provide your decision based on your profile and the instructions. Output *only* the JSON object.
\end{tcolorbox}

\begin{tcolorbox}[colback=blue!5!white, colframe=blue!75!black, breakable,title=Trustee Prompt (Trust Game)]
You are participating in a one-shot, anonymous economic game. Here is your profile: \{profile\_text\}\\

Experiment Instructions for You:

1. In this interaction, another anonymous participant started with \$10 and chose to send you an amount, N.

2. The amount they sent was tripled before you received it.

3. You have received a total of \$\{amount\_received\}. (This resulted from the other participant sending \$\{amount\_sent\}).

4. You now need to decide how much money, R, from the \$\{amount\_received\} you received, to send back to the first participant. R must be an integer between \$0 and \$\{amount\_received\}, inclusive.

5. This is a single interaction; you will not interact with this participant again.\\

Your Decision:

Based on your profile, the amount you received (\$\{amount\_received\}), and the instructions, decide the amount R (from 0 to \$\{amount\_received\}) you want to send back.\\

Output Format:

Provide your decision *only* as a JSON object with two keys:

- ``thinking\_process'': A brief explanation (1-2 sentences) of your reasoning for choosing the amount R, considering your profile, the amount received, and the situation.

- ``amount'': The integer amount R (0 to \$\{amount\_received\}) you decided to send back.\\

Example Output (if you received \$15):

\{\{

``thinking\_process'': ``They showed significant trust by sending \$5. It seems fair to return more than they sent, allowing both of us to profit from this interaction.'',

``amount'': 8

\}\}\\

Now, provide your decision based on your profile, the amount you received (\$\{amount\_received\}), and the instructions. Output *only* the JSON object.
\end{tcolorbox}

\begin{tcolorbox}[colback=blue!5!white, colframe=blue!75!black, breakable,title=Trustor Prompt (Dictator Game)]
You are participating in a one-shot, anonymous economic game. Here is your profile: \{profile\_text\}\\

Experiment Instructions for You:

1. You begin this interaction with \$10.

2. You need to decide how much money, N, to send to another anonymous participant. N must be an integer between \$0 and \$10, inclusive.

3. The amount N you choose to send will be tripled, and the other participant will receive \$3 * N.

4. The other participant cannot send any money back to you - they will simply receive whatever amount (3N) you decide to give them.

5. This is a single interaction; you will not interact with this participant again.\\

Your Decision:

Based on your profile and the instructions, decide the amount N (from 0 to 10) you want to send to the other participant (who will receive 3N).\\

Output Format:

Provide your decision *only* as a JSON object with two keys:

- ``thinking\_process'': A brief explanation (1-2 sentences) of your reasoning for choosing the amount N, considering your profile and the one-way nature of this interaction.

- ``amount'': The integer amount N (0-10) you decided to send.\\

Example Output:

\{\{

``thinking\_process'': ``I believe in helping others when I can, and knowing they'll receive triple what I send with no expectation of return, I'll send a modest amount.'',

``amount'': 3

\}\}\\

Now, provide your decision based on your profile and the instructions. Output *only* the JSON object.
\end{tcolorbox}

\begin{tcolorbox}[colback=blue!5!white, colframe=blue!75!black, breakable,title=Trustor Prompt (Repeated Trust Game)]
You are participating in a repeated economic game (7 rounds total). Here is your profile: \{profile\_text\}\\

\{history\_message\}\\

Experiment Instructions for You:

1. You begin this round with \$10.

2. You need to decide how much money, N, to send to the same participant you are paired with for all 7 rounds. N must be an integer between \$0 and \$10, inclusive.

3. The amount N you choose to send will be tripled, and the other participant will receive \$3 * N.

4. After receiving the tripled amount, the other participant will decide how much of that money (3N) to send back to you.

5. You will play 7 rounds total with the same participant, but each round starts fresh with \$10.\\

Your Decision:

Based on your profile and any history from previous rounds, decide the amount N (from 0 to 10) you want to send in this round.\\

Output Format:

Provide your decision *only* as a JSON object with two keys:

- ``thinking\_process'': A brief explanation (1-2 sentences) of your reasoning for choosing the amount N, considering your profile and the game history.

- ``amount'': The integer amount N (0-10) you decided to send.\\

Example Output:

\{\{

``thinking\_process'': ``Based on the previous round's interaction, I want to maintain trust while being slightly more cautious this time.'',

``amount'': 3

\}\}\\

Now, provide your decision based on your profile and the instructions. Output *only* the JSON object.
\end{tcolorbox}

\begin{tcolorbox}[colback=blue!5!white, colframe=blue!75!black, breakable,title=Trustee Prompt (Repeated Trust Game)]
You are participating in a repeated economic game (7 rounds total). Here is your profile: \{profile\_text\}\\

\{history\_message\}\\

Experiment Instructions for You:

1. In this round, the same participant you are paired with for all 7 rounds started with \$10 and chose to send you an amount, N.

2. The amount they sent was tripled before you received it.

3. You have received a total of \$\{amount\_received\}. (This resulted from the other participant sending \$\{amount\_sent\}).

4. You now need to decide how much money, R, from the \$\{amount\_received\} you received, to send back to the first participant. R must be an integer between \$0 and \$\{amount\_received\}, inclusive.

5. You will play 7 rounds total with the same participant, but each round starts fresh for the sender with \$10.\\

Your Decision:

Based on your profile, the amount you received (\$\{amount\_received\}), and any history from previous rounds, decide the amount R (from 0 to \$\{amount\_received\}) you want to send back.\\

Output Format:

Provide your decision *only* as a JSON object with two keys:

- ``thinking\_process'': A brief explanation (1-2 sentences) of your reasoning for choosing the amount R, considering your profile, the amount received, and the game history.

- ``amount'': The integer amount R (0 to \$\{amount\_received\}) you decided to send back.\\

Example Output (if you received \$15):

\{\{

``thinking\_process'': ``They maintained their trust level from last round. I'll reciprocate fairly to encourage continued cooperation.'',

``amount'': 9

\}\}\\

Now, provide your decision based on your profile, the amount you received (\$\{amount\_received\}), and the instructions. Output *only* the JSON object.
\end{tcolorbox}

\end{appendices}

\end{document}